\documentclass[12pt]{article}
\input epsf.tex
\usepackage{amsthm, amssymb, amsmath}

\begin{document}
\title{\bf{Ecological systems in a modeling perspective}}
\author{Torsten Lindstr\"{o}m \\
Department of Mathematics \\
Linnaeus University \\
SE-35195 V\"{a}xj\"{o}, SWEDEN\\
}
\date{}
\maketitle

\begin{abstract}
May (1974,1976)\nocite{May.Science:186,May.Nature:261} opened the debate on whether biological populations might exhibit nonlinear dynamics and chaos. However, it has in general been difficult to verify nonlinear dynamics in biological populations. There are many reports concerning problems with this issue and some of them can be traced back to Hassell, Lawton, and May (1976)\nocite{Hassel.JAE:45} and Morris (1990)\nocite{Morris.Ecology:71}.

Our objective is not a discussion of the presence of nonlinear dynamics in biological populations. Instead, we analyze whether ecological census data can be used for validating nonlinearities at all. We choose our models and our situation so that as much as possible can be done rigorously with by hand computations.

We consider a clearly nonlinear chemostat based model that is isolated. Some noise must be considered, and we choose a minimal approach: Only noise originating from the fact that ecological populations remain finite is considered, cf. Bailey (1964)\nocite{Bailey.Elements}.

Not only the interacting populations but also collected data sets tend to remain finite. Collection of long data sets might be associated with huge costs in ecology. Examples of exceptionally long and carefully studied ecological time series are those collected by Nicholson (1954)\nocite{Nicholson.AustJZool:2} and Utida (1957)\nocite{Utida.ColdSpring:22}. These data sets contain a few hundred data points, and we use this as a guideline for when an ecological time series should be considered exceptionally long in this chapter.
\end{abstract}

\section{Introduction}


Ecological systems differ in many aspects from those that we encounter in physics or chemistry. First, we might have serious difficulties in isolating the relevant parts of the system for a separate study. Second, the number of individuals operating in the system might be low or become low because of nonlinear oscillations calling usual modelling techniques like differential equation modelling into question. Under such circumstances, deterministic differential equations are replaced by a stochastic birth-death Markov chains (see e. g Bailey (1964)\nocite{Bailey.Elements}). The third problem is evolution, which limits our possibilities to repeat experiments under the same circumstances. The fourth problem is that information about mechanisms might exist, but information about them is simply not available in the data. That is, there are limits in using pattern detecting tools for detecting and formulating physical laws that are based on these mechanisms, cf. Lindstr\"{o}m (2009)\nocite{Lindstr.CSF:42}.

We consider a basic mechanistic model that is usually encountered in ecological modeling: the chemostat. In the chemostat a species is growing on a limiting nutrient. This assumption implies logistic growth of the species under certain assumptions. Chemostat ecologies have the benefit that the system can be kept isolated and that equations can be formulated for the processes that take place.

The logistic equation is nonlinear but separable. Thus, it allows for explicit solutions that can be fitted to census data. However, there is a limited number of cells in the culture, and the stochastic effects increase as the number of individuals in the population become smaller (Bailey (1964)\nocite{Bailey.Elements}). Stochastic versions of the logistic equation have been studied in detail by Renshaw (2011)\nocite{renshaw2}, but their parameters have not been explicitly linked to the parameters of the chemostat. Consequently, it remains invisible how the carrying capacity is linked to the resource dynamics.

Since the extinction state is absorbing, the stochastic logistic equation has no stationary distribution. However, conditioning on non-extinction gives a stationary solution and this distribution is called a quasi-stationary solution. If the quasi-stationary solution is approximately normal, then the expected time to extinction might be very long. In these cases, the quasi-stationary solution might describe the phenomenon that we are observing well. If the deterministic carrying capacity is sufficiently large, then the quasi-stationary population size becomes approximately normally distributed with mean at its deterministic carrying capacity, cf. N{\aa}sell (2001)\nocite{Naasell.JTB:211} and Renshaw (2011)\nocite{renshaw2}.

The logistic equation is, however, not the only equation that exhibits normally distributed stationary solutions or quasi-stationary solutions that could fit census data generated in chemostat situations. A standard technique in model selection is the use of methods that penalize the number of parameters in the model (including the intercept and the variance), see e.g. Burnham and Anderson (2002)\nocite{Burnham.Modelselect}.

Penalizing the number of parameters means penalizing contributions from second derivatives (Hastie and Tibshirani (1990)\nocite{Hastie}). That is, nonlinearities are penalized. Consequently, the correct model is penalized if the model that generated the data is nonlinear. The principle is that a nonlinear model including more parameters must produce a significantly better fit than a linear model to be justified from data. A convenient tool for doing this that has many good properties (Burnham and Anderson (2002)\nocite{Burnham.Modelselect}) is the Akaike (1973)\nocite{Akaike.second} Information Criterion (AIC). There are obviously many other ways to rank models with respect to data, too. A related method is the Bayesian Information Criterion (BIC), cf. Schwartz (1978)\nocite{Schwartz.AS:6}. Other examples are cross-validation and running-means (Hastie and Tibshirani (1990)\nocite{Hastie} and Green and Silverman (1994)\nocite{Green.Nonparam}).

Since the available data is usually limited in ecology, it sounds relevant to ask what data sets justify nonlinear models and in what way population size matter in this context. We want to keep the models that we use for generating our data as isolated as possible. Therefore, the natural noise that is generated by the assumption that we have a finite population size remains the only noise that we add to our models here.

The alternative models that we test with respect to our logistic chemostat data are linear on different scales and we assume that the data must support at least one additional parameter for justifying nonlinearity. The first model corresponds to a negative linear growth that is balanced by immigration. This does not imply that we have added any immigration to our chemostat - the model had been isolated. It has an approximately normal stationary distribution that shares expectation and variance of the normal distribution that approximates the quasi-stationary distribution of our logistic growth equation.

The second alternative model is linear on log-scale and equivalent to the Gompertz (1825)\nocite{Gompertz.TransLon:115} model. This model does not support any interpretation in terms of the modelled processes. It has an approximately normal stationary distribution that shares the expectation of the normal distribution that approximates the quasi-stationary distribution of our logistic growth equation. But its variance is larger than the variance of the approximate normal distribution of the quasi-stationary distribution of our logistic growth equation for high carrying capacities. It follows near normality (or high carrying capacities) that data-based criteria like the AIC (Akaike (1973)\nocite{Akaike.second}) and BIC (Schwartz (1978)\nocite{Schwartz.AS:6}) criteria rank this model below both the linear immigration model and the logistic growth equation that generated the data.

On the contrary, similar criteria rank the linear immigration model below both the Gompertz (1825)\nocite{Gompertz.TransLon:115} log-linear model and the logistic growth equation that generated the data for low carrying capacities far enough from normality. We conclude in Section \ref{detect_nonlin} that the ability of these methods to detect nonlinearities is dependent on the appearance of clusters of data corresponding to low population densities. Such clusters are expected to be rare when the quasi-stationary distribution is approximately normal. Indeed, the many of the most analyzed datasets in ecology (Nicholson (1954)\nocite{Nicholson.AustJZool:2} and Utida (1957)\nocite{Utida.ColdSpring:22}) remain at the limit for allowing a validation of possible nonlinearities with these criteria.

This chapter is structured as follows. We formulate our chemostat based model and derive the corresponding logistic model in Section \ref{modellen}. We derive discrete and stochastic variants of this model in Section \ref{Diskr_stok}. In Section \ref{appr_norm_sec} we compute the means and variances of the approximately normal quasi-stationary solution of our stochastic logistic model and compare it to real quasi stationary distributions for different availability of the limiting nutrient. In Section \ref{sec_mom_close} we use the moment closure method to receive different estimates of the mean and variance of the approximately normal quasi-stationary solution. The linear immigration comparison model is introduced in Section \ref{lin_immi} and our Gompertz (1825)\nocite{Gompertz.TransLon:115} log-linear model is introduced in Section \ref{log_linearity}. The corresponding differential equation is derived in Section \ref{log_linear_de}. Stochastic models corresponding to the linear immigration model and the Gompertz (1825)\nocite{Gompertz.TransLon:115} log-linear model are formulated and analyzed in Section \ref{Stochastic_models}. The possibilities for detecting non-linearities are discussed in Section \ref{detect_nonlin} and a summary of our results can be found in Section \ref{sum}. We support our claims for both low- and high carrying capacities with figures.

\section{The model}
\label{modellen}

We consider the chemostat system
\begin{eqnarray}
  \dot{\sigma} &=& CD-D\sigma-\frac{A\sigma\xi}{1+AB\sigma}, \nonumber\\
  \dot{\xi} &=& \frac{MA\sigma\xi}{1+AB\sigma}-D\xi.\label{chemo_density}
\end{eqnarray}
The variables, $\sigma$ and $\xi$, are non-negative and they stand for nutrient-con\-cen\-tra\-tion and the concentration of a species feeding on the nutrient. The parameters of the model are $C$, $D$, $A$, $B$, and $M$ are the nutrient-concentration in the inflow, the dilution-rate, the search-rate, the handling-time, and the conversion-factor for the species involved. The rational expressions used in (\ref{chemo_density}) corresponding to the specialist predator's functional response are justified by Holling (1959)\nocite{Holling.CanEnt:91}. The units and names of the parameters in (\ref{chemo_density}) are given in Table \ref{enhetstabell} together with their approximate magnitudes.
\begin{table}
\begin{tabular}{|l|l|l|l|}\hline
{\bf{Param.}} & {\bf{name}} & {\bf{unit}} & {\bf{Magn.}}\\\hline
$C$ & nutrient conc. in inflow & \# of particles/volume & $\sim 10^{23}$\\
$D$ & Dilution rate & fract. of cont./time& $\sim 10^{-1}$\\
$A$ & search rate & volume/time& $\sim 10^{-2}$\\
$B$ & handling time & time& $\sim 0$\\
$M$ & conversion factor & no unit& $\sim 10^{-23}$\\
$\sigma$ & nutrient conc. in cont & \# of particles/volume& $\sim 10^{23}$\\
$\xi$ & cell conc. in cont & \# of cells/volume & $\sim 10^1-10^3$\\
$L$ & volume of container & volume & $\sim10^0$\\
$S$ & nutrient part. in cont. & \# of particles & $\sim 10^{23}$\\
$X$ & cells in cont. & \# of cells &$\sim 10^1-10^3$\\
\hline
\end{tabular}
\caption{Parameters appearing in the models (\protect\ref{chemo_density})  and (\ref{chemo_abundance}) together with their units and approximate magnitudes.}
\label{enhetstabell}
\end{table}

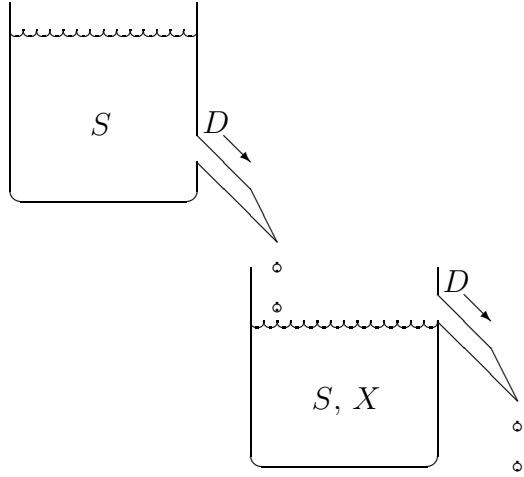
\begin{figure}
\begin{picture}(380,200)(0,0)
\put(122.5,170){\oval(5,5)[b]}
\put(127.5,170){\oval(5,5)[b]}
\put(132.5,170){\oval(5,5)[b]}
\put(137.5,170){\oval(5,5)[b]}
\put(142.5,170){\oval(5,5)[b]}
\put(147.5,170){\oval(5,5)[b]}
\put(152.5,170){\oval(5,5)[b]}
\put(157.5,170){\oval(5,5)[b]}
\put(162.5,170){\oval(5,5)[b]}
\put(167.5,170){\oval(5,5)[b]}
\put(172.5,170){\oval(5,5)[b]}
\put(177.5,170){\oval(5,5)[b]}
\put(182.5,170){\oval(5,5)[b]}
\put(187.5,170){\oval(5,5)[b]}

\put(120,180){\line(0,-1){70}}
\put(125,110){\oval(10,10)[bl]}
\put(125,105){\line(1,0){60}}
\put(185,110){\oval(10,10)[br]}
\put(190,110){\line(0,1){10}}
\put(190,120){\line(1,-1){30}}
\put(190,130){\line(1,-1){20}}
\put(210,110){\line(1,-2){10}}
\put(190,130){\line(0,1){50}}
\put(220,82){\circle*{1}}
\put(220,80){\circle{3}}
\put(220,79){\circle*{1}}
\put(150,130){$S$}

\put(192,131){$D$}
\put(200,130){\vector(1,-1){10}}
\put(220,67){\circle*{1}}
\put(220,65){\circle{3}}
\put(220,64){\circle*{1}}

\put(212.5,60){\oval(5,5)[b]}
\put(217.5,60){\oval(5,5)[b]}
\put(222.5,60){\oval(5,5)[b]}
\put(227.5,60){\oval(5,5)[b]}
\put(232.5,60){\oval(5,5)[b]}
\put(237.5,60){\oval(5,5)[b]}
\put(242.5,60){\oval(5,5)[b]}
\put(247.5,60){\oval(5,5)[b]}
\put(252.5,60){\oval(5,5)[b]}
\put(257.5,60){\oval(5,5)[b]}
\put(262.5,60){\oval(5,5)[b]}
\put(267.5,60){\oval(5,5)[b]}
\put(272.5,60){\oval(5,5)[b]}
\put(277.5,60){\oval(5,5)[b]}

\put(210,80){\line(0,-1){70}}
\put(215,10){\oval(10,10)[bl]}
\put(215,05){\line(1,0){60}}
\put(275,10){\oval(10,10)[br]}
\put(280,10){\line(0,1){50}}
\put(280,60){\line(1,-1){30}}
\put(280,70){\line(1,-1){20}}
\put(300,50){\line(1,-2){10}}
\put(280,70){\line(0,1){10}}
\put(310,22){\circle*{1}}
\put(310,20){\circle{3}}
\put(310,19){\circle*{1}}
\put(233,27){$S$, $X$}

\put(282,71){$D$}
\put(290,70){\vector(1,-1){10}}
\put(310,7){\circle*{1}}
\put(310,5){\circle{3}}
\put(310,4){\circle*{1}}
\end{picture}
\caption{Experimental setup for a chemostat. It is assumed that the fluids in both containers are kept well-mixed. This basic setup can be extended to contain more containers, species, and nutrients.}
\label{chemo_figure}
\end{figure}

If the size of the container is $L$, then we get with $S=L\sigma$, $X=L\xi$, $I_0=LCD$, that the abundance formulation of (\ref{chemo_density}) takes the form
\begin{eqnarray}
  \dot{S} &=&I_0-DS-\frac{ASX}{L+ABS}, \nonumber\\
  \dot{X} &=& \frac{MASX}{L+ABS}-DX,\label{chemo_abundance}
\end{eqnarray}
The experimental set-up corresponding to the equations (\ref{chemo_density}) and (\ref{chemo_abundance}) is given in figure \ref{chemo_figure}.
A standard technique for studying the dynamics of the chemostat is to study these equations in an asymptotically invariant manifold. We consider the functional
\begin{equation}
H(S,X)=MS+X-\frac{MI_0}{D}
\label{invariant-plane}
\end{equation}
and compute
\begin{eqnarray*}
\dot{H}&=&M\dot{S}+\dot{X}=\frac{DMI_0}{D}-MDS-DX\\
       &=&-D\left(MS+X-\frac{MI_0}{D}\right)=-DH.
\end{eqnarray*}
This proves that all solutions approach the manifold $H=0$ asymptotically. In this manifold we have
\begin{equation}
S=\frac{I_0}{D}-\frac{X}{M},
\label{invariant-plane_deriv_prop}
\end{equation}
so the system (\ref{chemo_abundance}) takes the form
\begin{displaymath}
 \dot{X} = \frac{MA\left(\frac{I_0}{D}-\frac{X}{M}\right)X}{L+AB\left(\frac{I_0}{D}-\frac{X}{M}\right)}-DX.
\end{displaymath}
With the assumption $B=0$ we get the logistic equation (Verhulst (1838)\nocite{verhulst1838})
\begin{displaymath}
  \dot{X} = M\frac{A}{L}\frac{I_0}{D}X-DX-\frac{A}{L}X^2=\left(M\frac{A}{L}\frac{I_0}{D}-D\right)X\left(1-\frac{X}{\frac{M\frac{A}{L}\frac{I_0}{D}-D}{\frac{A}{L}}}\right)
\end{displaymath}
with intrinsic growth rate
\begin{displaymath}
  R=M\frac{A}{L}\frac{I_0}{D}-D
\end{displaymath}
and carrying capacity
\begin{equation}
  K=\frac{M\frac{A}{L}\frac{I_0}{D}-D}{\frac{A}{L}}
  \label{Carrying}
\end{equation}
parameters. For the subsequent computations, we also introduce the birth-rate coefficient
\begin{equation}
  \rho=M\frac{A}{L}\frac{I_0}{D}\:\:{\rm{along \: with}}\:\:\alpha=\frac{A}{L}.
  \label{alpharho}
\end{equation}
We can then write $R=\rho-D$ and $K=R/\alpha$. We note that $K>0$ (or $R>0$) requires
\begin{displaymath}
M\frac{A}{L}\frac{I_0}{D}>D,
\end{displaymath}
i.e. the maximum specific growth rate exceeds the dilution rate. This provides a persistence condition for the deterministic chemostat treated so far here. We write
\begin{equation}
  \dot{X}=RX\left(1-\frac{X}{K}\right).
  \label{rk_logistic}
\end{equation}
We note that assigning meaningful interpretations to the parameters of the logistic equation is not straightforward, see e.g. Nisbet, McCauley, de Roos, Murdoch and Gurney (1991)\nocite{Nisbet.TPB:40}, Kuno (1991)\nocite{Kuno.RPE:33} and Kooi, Boer, and Kooijman (1998)\nocite{Kooi.BoMB:60}.

\begin{figure}
\epsfxsize=118mm
\begin{picture}(267,200)(0,0)
\put(45,30){\epsfbox{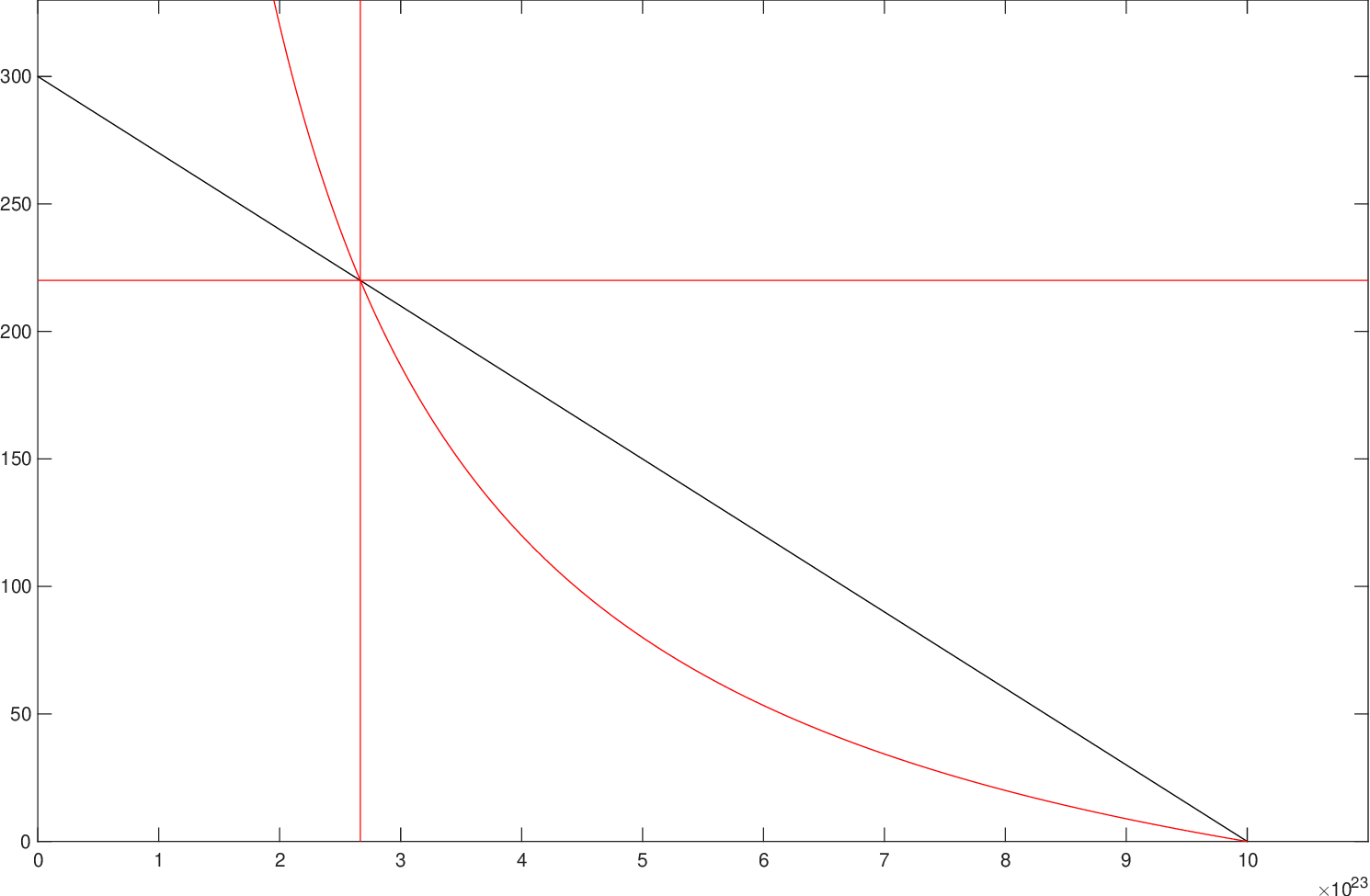}}
\put(5,228){$\frac{MI_0}{D}$}
\put(30,231){\vector(1,0){10}}
\put(0,178){$\frac{M\frac{A}{L}\frac{I_0}{D}-D}{\frac{A}{L}}$}
\put(45,181){\vector(1,0){10}}
\put(345,10){$\frac{I_0}{D}$}
\put(350,25){\vector(0,1){10}}
\put(125,10){$\frac{DL}{MA}$}
\put(134,25){\vector(0,1){10}}
\put(89,265){$X=\frac{L}{AS}(I_0-DS)$}
\put(108,265){\vector(1,-4){3}}
\put(124,206){\vector(0,-1){10}}
\put(116.5,230){\vector(0,-1){10}}
\put(133,235){\vector(-1,0){10}}
\put(133,215){\vector(-1,0){5}}
\put(134,145){\vector(1,0){5}}
\put(134,95){\vector(1,0){10}}
\put(180,112){\vector(0,1){10}}
\put(230,78){\vector(0,1){10}}
\put(182,150){\vector(-4,3){9}}
\put(260.5,100){\vector(-4,3){9}}
\put(88,210.5){\vector(4,-3){9}}
\put(60,43.5){\vector(1,0){10}}
\put(140,43.5){\vector(1,0){10}}
\put(220,43.5){\vector(1,0){10}}
\put(300,43.5){\vector(1,0){10}}
\put(54,210){\vector(2,-1){10}}
\put(54,150){\vector(2,-1){10}}
\put(54,90){\vector(2,-1){10}}
\put(95,150){\vector(3,-1){10}}
\put(95,90){\vector(4,-1){10}}
\put(160,85){\vector(3,2){10}}
\put(205,110){\vector(-2,3){7}}
\put(275,70){\vector(-1,1){10}}
\put(215,150){\vector(-2,1){10}}
\put(295,110){\vector(-2,1){10}}
\put(215,200){\vector(-3,1){10}}
\put(295,200){\vector(-3,1){10}}

\put(378,18){$S$}
\put(38,247){$X$}
\end{picture}
\caption{Sketch of the phase portrait of (\protect\ref{chemo_abundance}). The parameter values $M=3\cdot 10^{-22}$, $A=.01$, $B=0$, $I_0=8\cdot 10^{23}$, $D=.8$, and $L=1$ have been used. The expressions indicated for the different quantities in the figure are (if dependent on $B$) those corresponding to the case $B=0$. The slanted line is the asymptotically invariant manifold (\protect\ref{invariant-plane_deriv_prop}).}
\label{chemo_phase_portrait}
\end{figure}
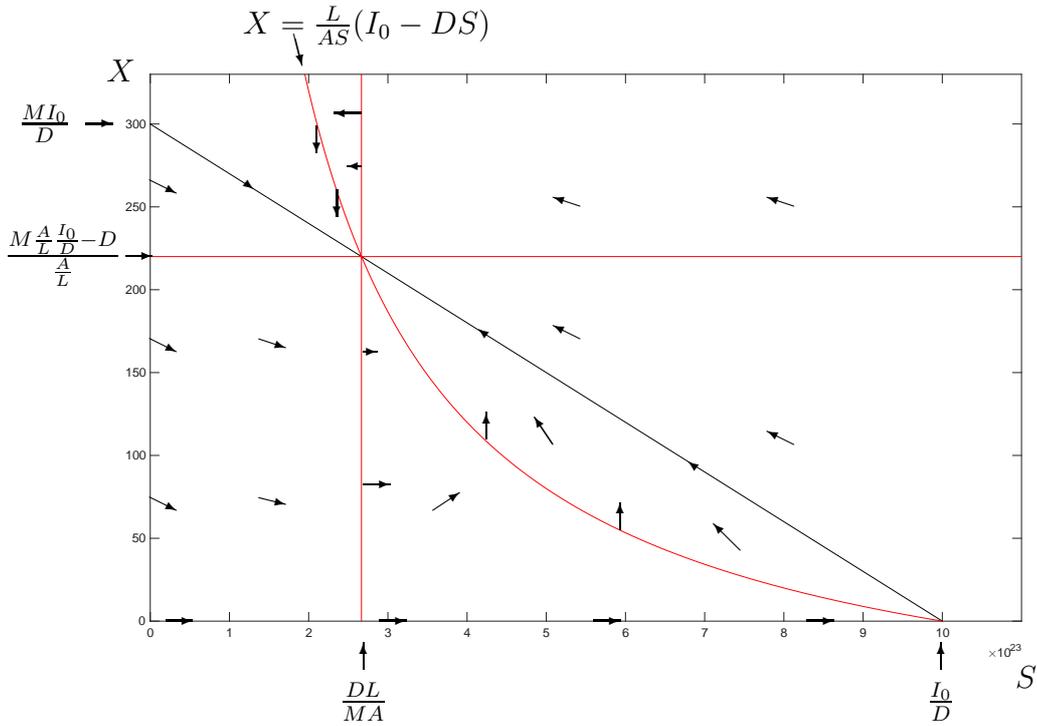

We sketch the major properties of the phase plot of (\ref{chemo_abundance}) in Figure \ref{chemo_phase_portrait} and it indicates an asymmetry between the parameters $M$ and $I_0$. The parameter $I_0$ has always large magnitude (order mols or $10^{23}$) whereas $M$ (and $B$) have small magnitudes (order $10^{-23}$). In the sequel, the parameters $M$ and $I_0$ always appear as a product (or order $1,\dots,10^3$). We are not going to use parameter $B$ here.

But setting $B=0$ does not mean that it wouldn't be important to find out e. g how large data sets are required for detecting saturation from data in Section \ref{detect_nonlin} or how the moment-closure method applies to saturated systems in Section \ref{sec_mom_close}. However, here we select carefully cases that allow as many steps as possible to be carried out explicitly with by hand calculations. Assuming $B\neq 0$ would introduce a difficult denominator already in the deterministic case and most likely, subsequent problems in our moment-closure analysis in Section \ref{sec_mom_close}. We remark that the explicit saturated deterministic (non-logistic) hyperbolic growth curve is available in Lindstr\"{o}m and Cheng (2015)\nocite{lindstr_cheng}.

\section{The connection to census data}
\label{Diskr_stok}

Data can be collected only at a limited number of time instances. This justifies a discrete variant of our deterministic model. The explicit solution of the logistic equation (\ref{rk_logistic}) can be computed by separation of variables. If the initial condition is given by $X_0$, then the solution $X(T)$ of (\ref{rk_logistic}) at time $T$ is given by
\begin{equation}
  X(T)=\frac{X_0\exp(RT)}{1+\frac{(\exp(RT)-1)X_0}{K}}=\frac{KX_0}{K\exp(-RT)+(1-\exp(-RT))X_0}.
  \label{diskret_logistic}
\end{equation}
We note that (\ref{diskret_logistic}) is the Beverton and Holt (1957)\nocite{Beverton.Dynamics} model and that this solution is a Poincar{\'{e}} map (Perko (2001)\nocite{perkobok_2001} and Hirsch, Smale and Devaney (2013)\nocite{hirschsmaledevaney}) of (\ref{rk_logistic}). Discrete population models are usually difficult to derive from mechanisms (cf. Lindstr\"{o}m (1999)\nocite{veszprem1}), but they represent the maps that are expected to describe the generation of census data. From (\ref{diskret_logistic}), we have $KX_0$ in the nominator and $K\exp(-RT)+(1-\exp(-RT))X_0$ in the denominator. Thus, the denominator interpolates between $X_0$ and $K$. For positive $X_0$ the denominator cannot become zero, so $X(T)$ starts from $X_0$ and proceeds with $T$ monotonically towards $K$.

Collection of data is not connected to a discrete collection of data only. It is connected to some noise, too. There are many ways to add noise to a model and the most natural way to add noise is to consider our model as an approximation of a Markov process. The model formulation in terms of abundances becomes now essential.

In a deterministic setting the univariate case takes the logistic form
\begin{equation}
  \dot{X} = M\frac{A}{L}\frac{I_0}{D}X-DX-\frac{A}{L}X^2= \rho X-DX-\alpha X^2.
  \label{logistic_det}
\end{equation}
if we use the parameters (\ref{alpharho}). The stochastic counterpart of (\ref{logistic_det}) is the birth-death Markov process (Bailey (1964)\nocite{Bailey.Elements} and Bhat and Miller (2002)\nocite{BhatMiller})
\begin{eqnarray}
\dot{p}_X&=&\rho(X-1)p_{X-1}-\rho Xp_X+D(X+1)p_{X+1}-DXp_X\nonumber\\
&&+\alpha(X+1)^2p_{X+1}-\alpha X^2p_X, \: X=1,2,3,\dots,\label{logistic_stok}\\
\dot{p}_0&=&\left(D+\alpha\right)p_1\nonumber
\end{eqnarray}
for the $X$-population which has been quite well-studied (Renshaw (2011)\nocite{renshaw2}). In (\ref{logistic_stok}), $p_X(t)$ stands for the probability that the population has size $X$ at time $t$. A stationary distribution corresponding to survival of the population does not exist in (\ref{logistic_stok}) since the extinction state is absorbing. However, conditioning on non-extinction, gives an approximately normal stationary solution whenever the growth rate is positive (N{\aa}sell (2001)\nocite{Naasell.JTB:211}). Stationary solutions that are obtained through conditioning on non-extinction are usually called quasi-stationary solutions.

\section{The diffusion approximation}
\label{appr_norm_sec}

There are some ways to approximate quasi-stationary solutions (cf. Renshaw (2011)\nocite{renshaw2}, p165-167). The diffusion approximation is based on the existence of sufficiently many individuals in the population (Allen (2008)\nocite{Brauer_Driesche_cpt3}) and we relate later in this section the phrase ``sufficiently many individuals'' to the parameters of our model. One way is to approximate (\ref{logistic_stok}) with its diffusion approximation
\begin{equation}
  \frac{\partial p_X}{\partial t}=-\frac{\partial\left(\left(\rho X-DX-\alpha X^2\right)p_X\right)}{\partial X}+\frac{1}{2}\frac{\partial^2\left(\left(\rho X+DX+\alpha X^2\right)p_X\right)}{\partial X^2}.\label{diff_appr}
\end{equation}
Equation (\ref{diff_appr}) may be written as a Fokker (1914)\nocite{Fokker.AdPh:348} and Plank (1917)\nocite{Planck} equation
\begin{equation}
  \frac{\partial p_X}{\partial t}=-\frac{\partial (f(X)p_X)}{\partial X}
  +\frac{1}{2}\frac{\partial^2 (g(X)p_X)}{\partial X^2}
  \label{diffusion_eq}
\end{equation}
with drift and diffusion functions represented as
\begin{eqnarray}
f(X)&=&\rho X-DX-\alpha X^2,\nonumber\\
g(X)&=&\rho X+DX+\alpha X^2.\label{logistic_fg}
\end{eqnarray}
Now, we write $\pi_X$ for the quasi-stationary distribution and get
\begin{displaymath}
  -\frac{(\partial f(X)\pi_X)}{\partial X}
  +\frac{1}{2}\frac{\partial^2 (g(X)\pi_X)}{\partial X^2}=0
\end{displaymath}
which after integration becomes
\begin{equation}
  -f(X)\pi_X+\frac{1}{2}\frac{\partial(g(X)\pi_X)}{\partial X}=-f(X)\pi_X
  +\frac{1}{2}g(X)\frac{\partial\pi_X}{\partial X}+\frac{1}{2}g^\prime(X)\pi_X=C_1.
  \label{integrated_eq}
\end{equation}
For large $X$, both the derivative and the value of the quasi-stationary distribution tend to be close to zero, so we may put $C_1=0$. Our objective is to approximate the quasi-stationary distribution in the vicinity of the deterministic carrying capacity given by (\ref{Carrying}).
We separate variables in (\ref{integrated_eq}) and get
\begin{displaymath}
 \frac{d(g(X)\pi_X)}{g(X)\pi_X}=2\frac{f(X)}{g(X)}dX
\end{displaymath}
which integrates into
\begin{equation}
\pi_X=\frac{C_2}{g(X)}\exp\left(2\int_{K}^{X}\frac{f(X^\prime)}{g(X^\prime)}dX^\prime\right)
\label{quasi_stat_distr}
\end{equation}
where the integration constant $C_2$ is selected to ensure that $\pi_X$ satisfies $\int_{0}^{\infty}\pi_X dX=1$. We estimate the integral for $X^\prime\approx K$ in (\ref{quasi_stat_distr}) by
\begin{eqnarray*}
  2\int_{K}^{X}\frac{f(X^\prime)}{g(X^\prime)}dX^\prime
&\approx&2\int_{K}^{X}\frac{f(K)+(X^\prime-K)\left.\frac{\partial f}{\partial X^\prime}\right|_{X^\prime=K}}{g(X^\prime)}dX^\prime\\
&\approx&2\int_{K}^{X}\frac{0+(X^\prime-K)\left.\frac{\partial f}{\partial X^\prime}\right|_{X^\prime=K}}{g(K)}dX^\prime\\
&=&\frac{(X-K)^2}{g(K)}\left.\frac{\partial f}{\partial X^\prime}\right|_{X^\prime=K}.
\end{eqnarray*}
We conclude that the diffusion approximation of the quasi-stationary distribution is a normal distribution with mean $K$ and variance
\begin{equation}
  V=-\frac{g(K)}{2\left.\frac{\partial f}{\partial X}\right|_{X=K}}
  \label{general_variance}
\end{equation}
in many cases. Thus, for computing the variance we need to compute both the drift derivative and the diffusion coefficient at $X=K$.

With the notation in (\ref{alpharho}) we get
\begin{displaymath}
  \left.\frac{\partial f}{\partial X}\right|_{X=K}=\rho-D-2\alpha K=R-2R=-R
\end{displaymath}
and
\begin{eqnarray*}
  g(K)&=&(\rho+D)K+\alpha K^2=(R+2D)\frac{R}{\alpha}+\frac{R^2}{\alpha}=\frac{R}{\alpha}(R+2D+R)\\
  &=&\frac{2R}{\alpha}(R+D)=\frac{2R\rho}{\alpha}.
\end{eqnarray*}

For (\ref{logistic_stok}) with positive carrying capacities, the diffusion approximation of the quasi-stationary distribution is approximately normal with expectation
\begin{equation}
  K=\frac{M\frac{A}{L}\frac{I_0}{D}-D}{\frac{A}{L}}
  \label{diffusionmean}
\end{equation}
(cf. (\ref{Carrying})) and variance
\begin{displaymath}
  V=-\frac{g(K)}{2\left.\frac{\partial f}{\partial X}\right|_{X=K}}=-\frac{\frac{2R\rho}{\alpha}}{-2R}=\frac{\rho}{\alpha}=\frac{M\frac{A}{L}\frac{I_0}{D}}{\frac{A}{L}}=\frac{MI_0}{D}.
\end{displaymath}

We return to the more general expression (\ref{general_variance}) for the diffusion approximated variance in Section \ref{Stochastic_models}. In Section \ref{Stochastic_models}, we also return to a representation of the variance
\begin{equation}
V=\frac{\rho}{\alpha}=\frac{R+D}{\alpha}=K+\frac{LD}{A}
\label{KdepVar}
\end{equation}
that expresses the variance as a function of the environmental carrying capacity. This expression can also be used for estimating the validity range of our diffusion approximations. Sufficiently large populations are usually quantified by small coefficients of variation. In terms of (\ref{KdepVar}) we get
\begin{equation}
{\rm{Coeff.\: of\: Variation}}=\frac{\sqrt{K+\frac{LD}{A}}}{K}<<1.
\label{Coeff_of_Variation}
\end{equation}
For $K>2$, the chain of inequalities
\begin{displaymath}
K(K-1)>K>\frac{LD}{A}
\end{displaymath}
implies that
\begin{equation}
K>>\frac{LD}{A}
\label{coeffvar}
\end{equation}
is a criterion in terms of the parameters of the model that ensures validity of our diffusion approximation.

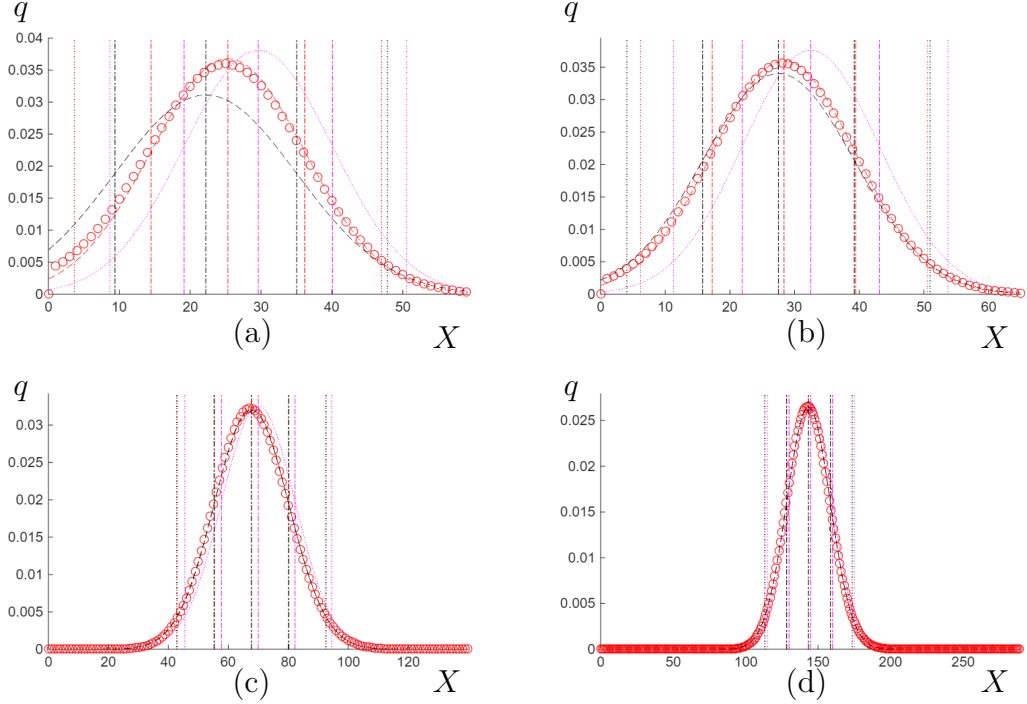
\begin{figure}
\epsfxsize=165mm
\begin{picture}(264,200)(0,0)
\put(-45,-15){\epsfbox{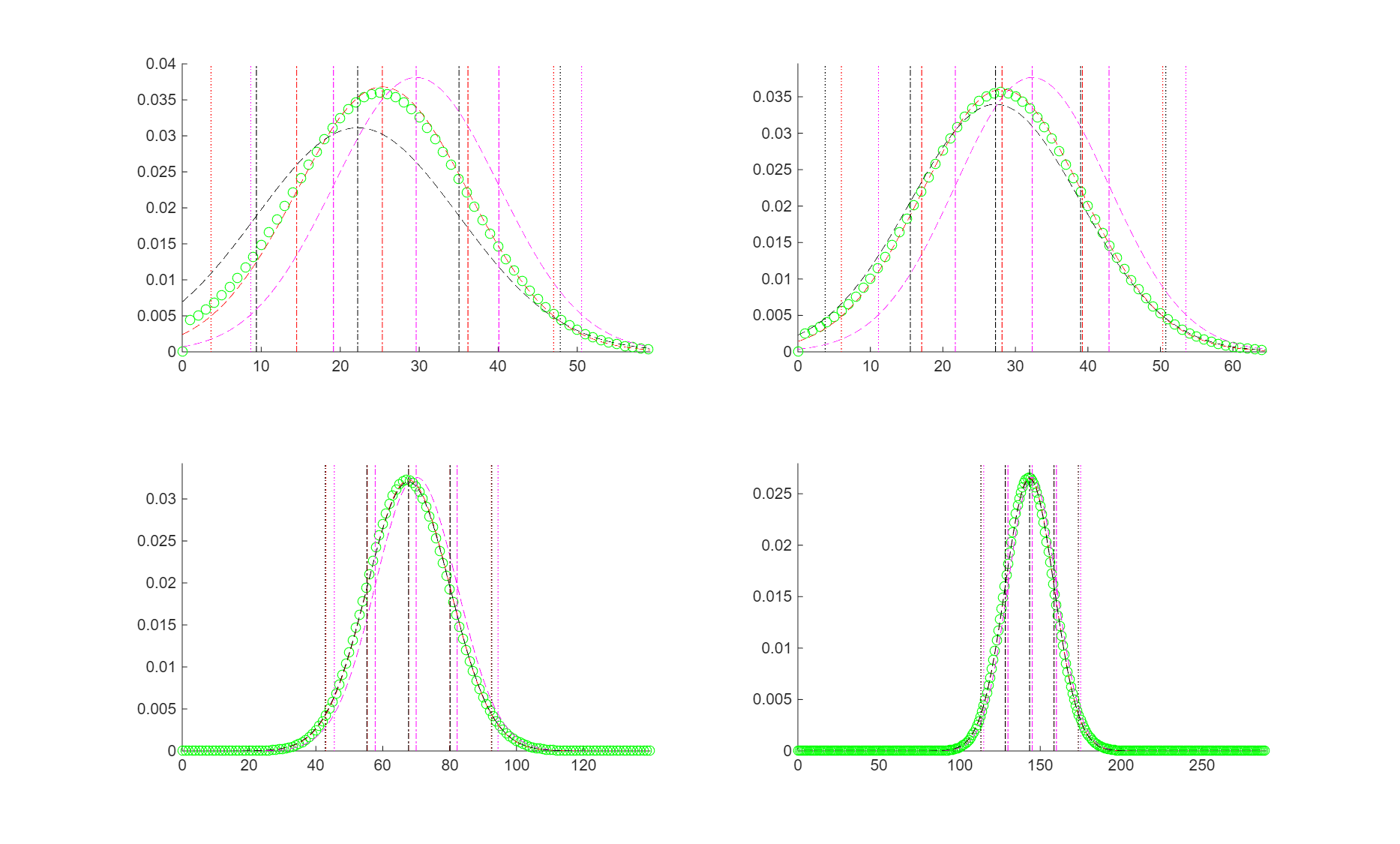}}

\put(160,130){$X$}
\put(3,255){$q$}
\put(85,132){(a)}

\put(365,130){$X$}
\put(209,255){$q$}
\put(292,132){(b)}

\put(160,0){$X$}
\put(3,112){$q$}
\put(85,0){(c)}

\put(365,0){$X$}
\put(209,112){$q$}
\put(292,0){(d)}
\end{picture}
\caption{Quasi-stationary probability mass functions of (\protect\ref{logistic_stok}) (green circles) compared to normal approximations sharing their expectations and variances (red dashed curves), to diffusion approximations (magenta dashed curves), and to moment closure approximations (black dashed curves). The parameter values are $M=3\cdot 10^{-22}$, $A=.01$, $D=.8$, $L=1$ and we have (a) $I_0=I_0^\ast(0)\approx 2.923 \cdot10^{23}$, (b) $I_0\approx 2.995\cdot10^{23}$, (c) $I_0=4\cdot10^{23}$, and (d) $I_0=6\cdot10^{23}$. The red lines correspond to the expectations, expectations plus/minus one standard deviation, and to expectations plus/minus two standard deviations. The magenta lines correspond to the expectations and standard deviations of the diffusion approximation. The black lines correspond to the expectations and standard deviations of the moment closure approximation. The selection of the parameter value in (b) will be explained in Section \protect\ref{Stochastic_models}.}
\label{normal_distr_appr1}
\end{figure}

A picture of how a normal distribution approximates the actual distribution as the carrying capacity becomes larger and the variance becomes relatively smaller is given in Figure \ref{normal_distr_appr1}(a)-(d). Here, we have depicted the actual distribution by green circles, a normal distribution that shares the mean and variance with the actual distribution with a red dashed curve, and the approximate normal distribution derived above with a magenta curve. In Figure \ref{normal_distr_appr1}(a)-(d), we have
\begin{displaymath}
  \frac{LD}{A}=80.
\end{displaymath}
Thus, criterion (\ref{coeffvar}) implies that panel (c) is at the validity boundary and that panel (d) has a clearly valid diffusion approximation whereas panels (a) and (b) remain quite far from validity. We have added the normal distribution curve that corresponds to moment closure approximation (Section \ref{sec_mom_close}) as a black dashed curve in the panels (a)-(d) and we note that this approximation approaches the actual probability mass function visibly faster than the diffusion approximation in panels (b) and (c). However, the precise expressions for the variances and means are far more complicated in this case.

We conclude that the risk of extinction for smaller populations with comparably large variances is considerably higher than for larger populations with comparably small variances. The concept of quasi-stationarity receives its meaning in Figure \ref{normal_distr_appr1}(c) when the probabilities for populations sizes in the span $1,\dots,30$ become extremely small because of approximate normality of the distribution. There is almost no probability that the population size passes this interval given that it has already reached a value close to the carrying capacity.

\section{The moment closure approximation}
\label{sec_mom_close}

The fact that the quasi-stationary distribution is approximately normal opens another possibility to analyze the properties of this distribution, too. These approximations have been frequently used in epidemiology and are usually called moment closure approximations (Lloyd (2004)\nocite{Lloyd.TPB:65}, Marrec, Bank, and Bertrand (2023)\nocite{Marrec.EcolEvol:2}, and Trostle, Guinness, and Reich (2024)\nocite{Trostle.Biometrics:80}). They add understanding to both the deterministic approximation and the diffusion approximation. The technique itself can be traced back to Whittle (1957)\nocite{Whittle.JRSS_B:19}. It has, since then been improved step by step, see e.g. N{\aa}sell (2003)\nocite{Naasell.TPB:64} and N{\aa}sell (2017)\nocite{Naasell.BoMB:79}.

We consider the probability distribution $q_X(t)$, $X=1,2,\dots$ defined by the conditional probabilities
\begin{equation}
  q_X(t)=\frac{p_X(t)}{1-p_0(t)}.
  \label{def_quasi}
\end{equation}
of the probabilities in (\ref{logistic_stok}). If we combine (\ref{logistic_stok}) and (\ref{def_quasi}) we get
\begin{eqnarray}
\dot{q}_X&=&\rho(X-1)q_{X-1}-\rho Xq_X+D(X+1)q_{X+1}-DXq_X+\alpha(X+1)^2q_{X+1}\nonumber\\
&&-\alpha X^2q_X+Dq_1q_X+\alpha q_1q_X, \: X=1,2,\dots,\label{logistic_stok_quasi}
\end{eqnarray}
with $q_0=0$. The stationary distribution of (\ref{logistic_stok_quasi}) is the quasi-stationary distribution of (\ref{logistic_stok}).
We expect that this distribution is approximately normal and we use the fact that the normal distribution is the only distribution for which the only non-zero cumulants are the mean and the variance. The moment generating function with respect to our quasi-stationary distribution $q_X$ is defined as
\begin{displaymath}
{\cal{M}}(\theta,t)=E[e^{\theta X}]=\sum_{X=1}^\infty e^{\theta X}q_X(t).
\end{displaymath}
We can derive a partial differential equation for the moments of the time-dependent probability distribution $q_X(t),\: X=1,2\dots,t\geq 0$ if we multiply the equations in (\ref{logistic_stok_quasi}) with $e^{\theta X}$, sum them, and then use the identities
\begin{equation}
  \frac{\partial {\cal{M}}}{\partial t}=\sum_{X=1}^\infty e^{\theta X}\dot{q}_X(t),\:\:
  \frac{\partial {\cal{M}}}{\partial \theta}=\sum_{X=1}^\infty Xe^{\theta X}q_X(t),\:\:{\rm{and}}\:\:
  \frac{\partial^2 {\cal{M}}}{\partial \theta^2}=\sum_{X=1}^\infty X^2e^{\theta X}q_X(t).
  \label{partial_diff_equal}
\end{equation}
We get
\begin{eqnarray*}
\sum_{X=1}^\infty e^{\theta X}\dot{q}_X(t)&=&\sum_{X=1}^\infty \rho e^{\theta X}(X-1)q_{X-1}-\sum_{X=1}^\infty \rho e^{\theta X}Xq_X\\
&&+\sum_{X=1}^\infty e^{\theta X}D(X+1)q_{X+1}-\sum_{X=1}^\infty e^{\theta X}DXq_X\\&&+\sum_{X=1}^\infty\alpha e^{\theta X}(X+1)^2q_{X+1}-\sum_{X=1}^\infty\alpha e^{\theta X}X^2q_X\\
&&+Dq_1\sum_{X=1}^\infty e^{\theta X}q_X+\alpha q_1\sum_{X=1}^\infty e^{\theta X}q_X\\
&=&\rho\sum_{X=0}^\infty e^{\theta (X+1)}Xq_{X}-\rho\sum_{X=1}^\infty e^{\theta X}Xq_X+D\sum_{X=1}^\infty e^{\theta (X-1)}Xq_{X}\\
&&-D\sum_{X=1}^\infty e^{\theta X}Xq_X+\alpha\sum_{X=1}^\infty e^{\theta (X-1)}X^2q_{X}-\alpha\sum_{X=1}^\infty e^{\theta X}X^2q_X\\
&&+Dq_1\sum_{X=1}^\infty e^{\theta X}q_X+\alpha q_1\sum_{X=1}^\infty e^{\theta X}q_X-Dq_1-\alpha q_1.
\end{eqnarray*}
We use the equalities (\ref{partial_diff_equal}) to obtain the partial differential equation
\begin{eqnarray}
  \frac{\partial {\cal{M}}}{\partial t}&=&\rho(e^\theta-1)\frac{\partial {\cal{M}}}{\partial\theta}+D(e^{-\theta}-1)\frac{\partial {\cal{M}}}{\partial\theta}
  +\alpha(e^{-\theta}-1)\frac{\partial^2 {\cal{M}}}{\partial\theta^2}\nonumber\\
  &&+q_1\left(D+\alpha\right)({\cal{M}}-1)
  \label{Moment_PDE}
\end{eqnarray}
for the moment generating function. The cumulant generating function is defined by ${\cal{K}}(\theta,t)=\log{\cal{M}}(\theta,t)$. We get the corresponding partial differential equation for the cumulant generating function if we substitute the transformation identities
\begin{displaymath}
  \frac{1}{{\cal{M}}}\frac{\partial {\cal{M}}}{\partial t}=\frac{\partial {\cal{K}}}{\partial t},
\:\frac{1}{{\cal{M}}}\frac{\partial {\cal{M}}}{\partial\theta}=\frac{\partial {\cal{K}}}{\partial\theta},
\:\frac{1}{{\cal{M}}}\frac{\partial^2{\cal{M}}}{\partial\theta^2}=\left(\frac{\partial {\cal{K}}}{\partial\theta}\right)^2+\frac{\partial^2 {\cal{K}}}{\partial\theta^2}
\end{displaymath}
into (\ref{Moment_PDE}). The resulting partial differential equation for the cumulant generating function is thus, given by
\begin{eqnarray}
  \frac{\partial {\cal{K}}}{\partial t}&=&\rho(e^\theta-1)\frac{\partial {\cal{K}}}{\partial\theta}+D(e^{-\theta}-1)\frac{\partial {\cal{K}}}{\partial\theta}\nonumber\\
  &&+\alpha(e^{-\theta}-1)\left(\left(\frac{\partial {\cal{K}}}{\partial\theta}\right)^2+\frac{\partial^2 {\cal{K}}}{\partial\theta^2}\right)
  +q_1\left(D+\alpha\right)\left(\frac{e^{\cal{K}}-1}{e^{\cal{K}}}\right).
  \label{Kumulat_PDE}
\end{eqnarray}
A series expansion of the cumulant generating function now gives
\begin{eqnarray*}
  {\cal{K}}(\theta,t)&=&E[X(t)]\theta+\frac{1}{2!}V[X(t)]\theta^2+\dots+{\rm{higher\: order\: cumulants,}}\\
  \frac{\partial{\cal{K}}}{\partial t}(\theta,t)&=&\dot{E}[X(t)]\theta+\frac{1}{2!}\dot{V}[X(t)]\theta^2+\dots+{\rm{higher\: order\: cumulants,}}\\
  \frac{\partial{\cal{K}}}{\partial\theta}(\theta,t)&=&E[X(t)]+V[X(t)]\theta+\dots+{\rm{higher\: order\: cumulants,}}\\
  \frac{\partial^2{\cal{K}}}{\partial\theta^2}(\theta,t)&=&V[X(t)]+\dots+{\rm{higher\: order\: cumulants,}}\\
  \frac{e^{\cal{K}}-1}{e^{\cal{K}}}&=&E[X(t)]\theta+\frac{1}{2!}(V[X(t)]+E^2[X(t)])\theta^2+\dots+{\rm{h.\: o.\: cmlts.}}
\end{eqnarray*}
If we substitute these expressions into (\ref{Kumulat_PDE}) we get infinitely many differential equations. We expect that the first two of them with higher order cumulants equated to zero works close to the approximately normal quasi-stationary distribution. The equations read
\begin{eqnarray}
  \dot{E}[X] &=& \rho E[X]-DE[X]-\alpha E^2[X]-\alpha V[X]+q_1\left(D+\alpha\right)E[X], \nonumber\\
  \dot{V}[X] &=& \rho E[X]+DE[X]+\alpha E^2[X]+\alpha V[X]-2DV[X]\label{Moment_close_ode} \\
  && -4\alpha E[X]V[X]+2\rho V[X]+q_1\left(D+\alpha\right)\left(E^2[X]+V[X]\right).\nonumber
\end{eqnarray}
The isoclines of (\ref{Moment_close_ode}) consist of a hyperbola (from the second equation) and a parabola (from the first equation). They always intersect at $E[X]=0$ and $V[X]=0$. Consequently, the origin is always an equilibrium. It is unstable by the trace-determinant criterion for parameter values of interest here. The solutions of a quadratic equation give the remaining possible equilibria. A saddle-node bifurcation occurs at
\begin{equation}
  I_0^\ast(q_1)=\frac{D}{M}\left(4+D\frac{L}{A}+2\sqrt{4+2D\frac{L}{A}+q_1\left(1+D\frac{L}{A}\right)}\right)
\label{I0q1}
\end{equation}
expressed in the original chemostat parameters.
The computations ending in this threshold proceed through equalities and do not include difficult estimates. They can probably easily be checked or followed with some computer algebra system. We included some of the details in Appendix \ref{saddlenodethresholdchapter} for convenience.

For values with $I_0>I_0^\ast(q_1)$ the moment closure approximation has two additional equilibria at
\begin{eqnarray}
E[X]&=&\frac{3R+2q_1\left(D+\alpha\right)\pm\sqrt{R^2-4q_1\alpha\left(D+\alpha\right)-8\alpha\rho}}{4\alpha},\nonumber\\
V[X]&=&\frac{\alpha\rho}{2\alpha}+\frac{R^2}{8\alpha}+\frac{q_1\left(D+\alpha\right)}{4\alpha}\left(\alpha+q_1\left(D+\alpha\right)+2R\right)\label{mom_closure_est}\\
&&\mp\frac{R\sqrt{R^2-4q_1\alpha\left(D+\alpha\right)-8\alpha\rho}}{8\alpha}.\nonumber
\end{eqnarray}
In (\ref{mom_closure_est}), the equilibrium corresponding to the sign below the other is a saddle and the equilibrium corresponding to the sign above the other is stable. Stability follows since we have
\begin{displaymath}
  \frac{3R+2q_1\left(D+\alpha\right)+\sqrt{R^2-4q_1\alpha\left(D+\alpha\right)-8\alpha\rho}}{4\alpha}>\frac{3}{4}K+\frac{q_1}{2}\left( 1+\frac{D}{\alpha}\right)
\end{displaymath}
and thus, the trace of the Jacobian can be estimated as
\begin{eqnarray*}
\frac{{\rm{Tr}}J(E[X],V[X])}{\frac{A}{L}}&=&3K+1-6E[X]+2q_1\left(1+\frac{D}{\alpha}\right)\\
&\leq&3K+1-6\cdot\frac{3}{4}K-3q_1\left(1+\frac{D}{\alpha}\right)+2q_1\left(1+\frac{D}{\alpha}\right)\\
&\leq&-\frac{3}{2}K-q_1\left(1+\frac{D}{\alpha}\right)+1<0.
\end{eqnarray*}
The first inequality follows from (\ref{mom_closure_est}) and the last inequality is based on that the carrying capacity must exceed $\frac{2}{3}$ individuals. It follows that every viable population consisting of at least one individual must meet this condition. Index theory (Jordan and Smith (1990)\nocite{jordan}, Dumortier (2006)\nocite{Dumortierbook}) ensures that the index of all fixed points in the non-negative quadrant equals one. The location of the stable fixed point is the moment-closure approximation. The stable manifold of the saddle point bounds its domain of attraction.

We estimate $q_1$ in (\ref{I0q1}) and denote the probability distribution of the standard normal distribution by
\begin{displaymath}
  \psi(x)=\frac{1}{\sqrt{2\pi}}e^{-\frac{x^2}{2}}.
\end{displaymath}
A normal approximation with mean $K$ cf. (\ref{diffusionmean}) and variance $V=K+\frac{LD}{A}$ cf. (\ref{KdepVar}) gives now
\begin{displaymath}
  q_1\approx\frac{\psi(\frac{-K+1}{\sqrt{V}})}{\sqrt{V}}<<1
\end{displaymath}
when $K>>\sqrt{V}\approx\sqrt{K}$ or $K>>1$, cf. (\ref{Coeff_of_Variation}). It follows that $q_1\approx 0$. This approximation improves as the normal approximation achieves more precision. We use $I_0^\ast(0)$ in our numerical work below.

\begin{figure}
\epsfxsize=137mm
\begin{picture}(267,200)(0,0)
\put(0,15){\epsfbox{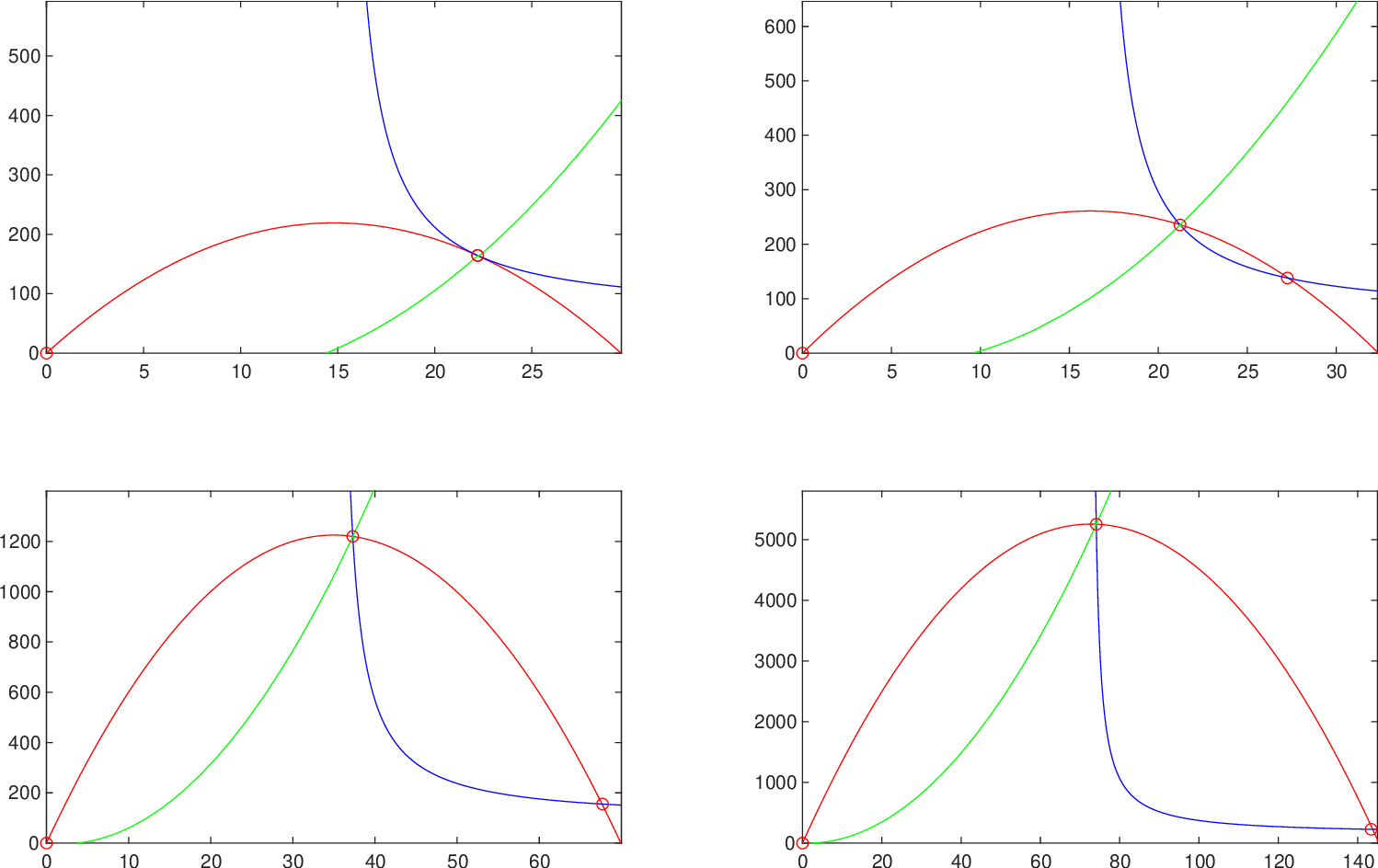}}
\put(85,5){(c)}
\put(146,5){$E[X]$}
\put(300,5){(d)}
\put(363,5){$E[X]$}
\put(-15,120){$V[X]$}
\put(-15,260){$V[X]$}
\put(85,143){(a)}
\put(146,143){$E[X]$}
\put(300,143){(b)}
\put(363,143){$E[X]$}
\put(199,120){$V[X]$}
\put(199,260){$V[X]$}

\put(35,225){\vector(-2,3){7}}
\put(160,235){\vector(-1,-1){10}}
\put(104,245){\vector(-1,0){10}}
\put(107.5,225){\vector(-1,0){10}}
\put(115,205){\vector(-1,0){10}}
\put(165,180){\vector(-1,0){10}}
\put(170,166){\vector(0,1){10}}
\put(70,194){\vector(0,1){10}}
\put(40,181.5){\vector(0,1){10}}
\put(157.5,210){\vector(-1,-1){5}}
\put(107,168.5){\vector(2,1){5}}
\put(80,175){\vector(2,3){7}}
\put(70,235){\vector(-3,2){10}}
\put(135,235){\vector(-1,-1){10}}
\put(140,165){\vector(2,3){7}}
\put(165,200){\vector(-1,-1){10}}

\put(340,230){\vector(-1,-1){10}}
\put(380,205){\vector(-1,-1){10}}
\put(317,246){\vector(-1,0){10}}
\put(323,215){\vector(-1,0){10}}
\put(341,190.5){\vector(1,0){7}}
\put(383,179){\vector(-1,0){10}}
\put(350,190){\vector(0,-1){5}}
\put(384.5,166){\vector(0,1){10}}
\put(280,196){\vector(0,1){10}}
\put(250,181){\vector(0,1){10}}
\put(361,228){\vector(-2,-3){5}}
\put(300,171){\vector(2,1){5}}
\put(260,215){\vector(-2,3){7}}
\put(285,185){\vector(3,2){10}}
\put(335,170){\vector(1,1){10}}

\put(35,90){\vector(-2,3){7}}
\put(160,110){\vector(-1,-1){10}}
\put(103,80){\vector(1,0){10}}
\put(112,50){\vector(1,0){10}}
\put(140,36.5){\vector(1,0){10}}
\put(125,96){\vector(0,-1){10}}
\put(150,69){\vector(0,-1){10}}
\put(48,82){\vector(0,1){10}}
\put(25,46){\vector(0,1){10}}
\put(75,65){\vector(2,3){5}}
\put(40,28.5){\vector(2,1){5}}
\put(104,118){\vector(-1,-2){2}}
\put(45,50){\vector(2,3){7}}
\put(80,40){\vector(3,2){10}}
\put(130,60){\vector(1,-1){10}}

\put(245,80){\vector(-2,3){7}}
\put(380,90){\vector(-1,-1){10}}
\put(310,117){\vector(-1,0){10}}
\put(311,85){\vector(1,0){10}}
\put(314,52){\vector(1,0){10}}
\put(340,28.5){\vector(1,0){10}}
\put(346,92){\vector(0,-1){10}}
\put(369,62){\vector(0,-1){10}}
\put(271.5,94){\vector(0,1){10}}
\put(238,45){\vector(0,1){10}}
\put(284.5,65){\vector(2,3){5}}
\put(255,31){\vector(1,1){5}}
\put(314,121){\vector(-1,-2){2}}
\put(260,50){\vector(2,3){7}}
\put(285,40){\vector(3,2){10}}
\put(340,60){\vector(1,-1){10}}

\end{picture}
\caption{Phase portraits of (\protect\ref{Moment_close_ode}). The parameter values are $M=3\cdot 10^{-22}$, $A=.01$, $D=.8$, $L=1$, $q_1=0$ and we have (a) $I_0=I_0^\ast(0)\approx 2.923 \cdot10^{23}$, (b) $I_0\approx 2.995\cdot10^{23}$, (c) $I_0=4\cdot10^{23}$, and (d) $I_0=6\cdot10^{23}$. The blue and the red curves are the isoclines of (\protect\ref{Moment_close_ode}) and the green curve is the stable manifold of the saddle point given by (\protect\ref{mom_closure_est}). The saddle point and the stable fixed point collide and disappear at the saddle-node-bifurcation at $I_0=I_0^\ast(0)\approx 2.923 \cdot10^{23}$. The selection of the parameter value in (b) will be explained in Section \protect\ref{Stochastic_models}.}
\label{Moment_phase_portrait}
\end{figure}
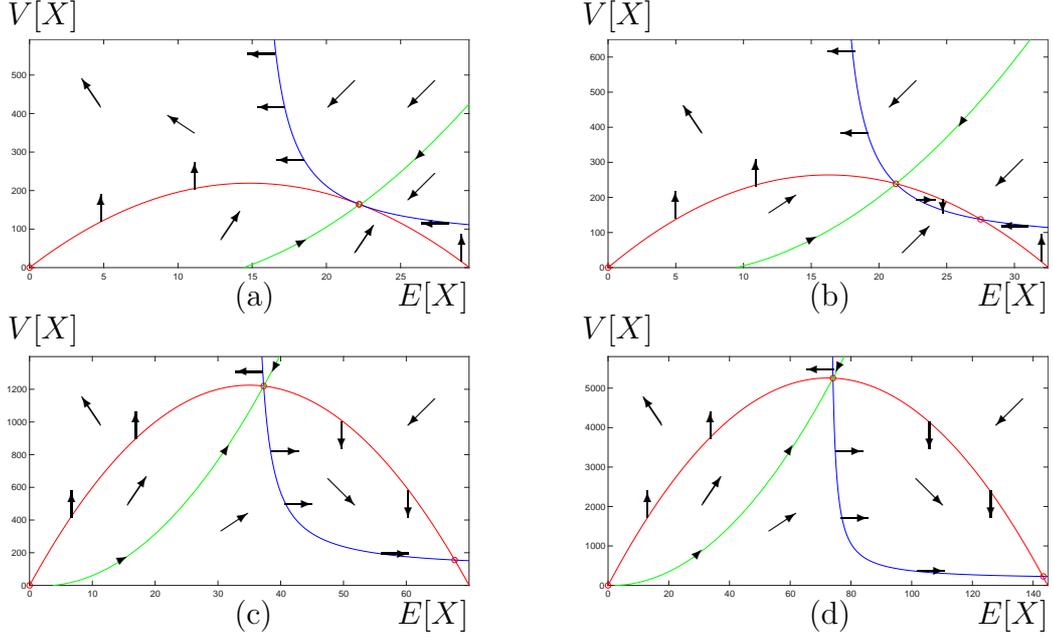

We have depicted the phase-plane of (\ref{Moment_close_ode}) in Figure \ref{Moment_phase_portrait}. The parabola isocline originating from the first equation of (\ref{Moment_close_ode}) is depicted by a red curve and the hyperbola originating from the second equation is depicted by a blue curve. The green curve is the stable manifold of the saddle point given by (\ref{mom_closure_est}). The part below the green curve is the attraction domain of the stable fixed point given by (\ref{mom_closure_est}).

That the moment closure method leads to the emergence of additional spurious equilibria (like the origin and the saddle point here) was already noted by Whittle (1957)\nocite{Whittle.JRSS_B:19}. The attraction domain of the stable equilibrium depends on how the moment closure method is applied (N{\aa}sell (2003)\nocite{Naasell.TPB:64}). Indeed, the quasi-stationary distribution can be approximated with several other distributions and the outcome of approximating it with e.g. binomial or log-normal distributions influences the attraction domain. In particular, the attraction domain might include the whole non-negative cone for good choices of the approximation.

Taken together, normal approximations arise from the moment closure method when all cumulants except for those of order one and two are put to zero. Similarly, the deterministic approximation arises when all cumulants except for those of order one are put to zero. The deterministic approximation is, however, safer in the sense that the equations are easier to check and that spurious fixed points with corresponding divisions in safe and non-safe parts of the phase space do not arise. As we have seen here, one of drawbacks is that this method easily leads to complicated equations and computations. Another drawback is that it is not entirely clear when and why the method works (cf. Kuehn (2016)\nocite{Kuehn_Mom_closure}).

We see that the results presented here are not very dependent on the small parameter $q_1(t)$ that originates from the fact that we are dealing with the quasi-stationary distribution $q$ and not with the original distribution $p$. The asymptotic properties of the cumulants computed from these distributions differ since all the cumulants of the original distribution $p$ tend to zero because of the absorbing state at $X=0$. Therefore, one should use the cumulants computed with respect to the quasi-stationary distribution that contains the properties of the approximate normal distribution (N{\aa}sell (2017)\nocite{Naasell.BoMB:79}).

\section{A linear immigration model}
\label{lin_immi}

Our next question is under what circumstances there are possibilities to identify (\ref{diskret_logistic}) from data generated by the birth-death process (\ref{logistic_stok}). It follows from uniqueness of solutions of (\ref{rk_logistic}) that its Poincar{\'{e}} map (\ref{diskret_logistic}) is strictly increasing with respect to $X_0$, cf. (\ref{the_slope}). Thus, (\ref{diskret_logistic}) cannot possess nonlinear dynamics like limit cycles or chaos. Our focus becomes detecting the nonlinearity in the map itself.

For accomplishing this, we start our analysis from the solutions of the differential equation
\begin{equation}
  \dot{X}=a+bX+cX^2.
  \label{quadr_model}
\end{equation}
We note that (\ref{logistic_det}) is (\ref{quadr_model}) with $a=0$, $b=R$, and $c=-\frac{R}{K}$.
The model (\ref{quadr_model}) is linear when $c=0$, and then it possesses the solution
\begin{equation}
X(T)=\frac{a}{b}(\exp(bT)-1)+\exp(bT)X_0.
\label{Poinc_lin}
\end{equation}
The Poincar{\'{e}} map (\ref{Poinc_lin}) of the linear part of (\ref{quadr_model}) is linear, too. It has slope $\exp(bT)$. The qualitative dynamical behavior of (\ref{Poinc_lin}) corresponds to a globally stable fixed point at $X_0=-\frac{a}{b}$ if $b<0$. This fixed point is non-negative, if $a\geq 0$.

Now we are returning to the model (\ref{diskret_logistic}). For positive values of $X_0$, it has a globally stable fixed point at $X_0=K$. Its slope is given by
\begin{equation}
  \frac{dX(T)}{dX_0}=\frac{K^2e^{-RT}}{(K\exp(-RT)+(1-\exp(-RT))X_0)^2}.
  \label{the_slope}
\end{equation}
Evaluated at $X_0=K$, this takes the value
\begin{displaymath}
  \left.\frac{dX(T)}{dX_0}\right|_{X_0=K}=\exp(-RT).
\end{displaymath}
Requiring that the location and the stability properties of the fixed points of the models (\ref{diskret_logistic}) and (\ref{Poinc_lin}) agree provides the linear system
\begin{eqnarray*}
  K &=& -\frac{a}{b}, \\
  \exp(-RT) &=&\exp(bT).
\end{eqnarray*}
Back-substitution gives $b=-R$ and $a=KR$. A linearization of (\ref{diskret_logistic}) is therefore, given by
\begin{equation}
  X(T)=K(1-\exp(-RT))+\exp(-RT)X_0.
  \label{diskret_linear}
\end{equation}
This linearization has a fixed point at $X(T)=X_0$ and has slope $\exp(-RT)$ at that fixed point. The differential equation of form (\ref{quadr_model}) that corresponds to this discrete model is
\begin{equation}
  \dot{X}=R(K-X).
  \label{lin_de}
\end{equation}
It corresponds to a negative growth rate that is balanced by a positive immigration rate. It is a possible local population model, but the mechanisms included here can never correspond to a global population model. We note that assumptions very similar to the assumptions made in the model (\ref{lin_de}) are commonly used in constant recruitment rate models in epidemiology, see e.g. Saha and Ghosh (2023)\nocite{Saha.IJDC:11} and references therein.

\section{A log-scale linearization}
\label{log_linearity}

Linearization without change of scale resulted in a model that replaced the logistically controlled population growth by a negative growth compensated by migration. This method could not yield any valid mechanisms for a global population model. Another way to linearize population models is to do the linearization on log-scale, cf. Gompertz (1825)\nocite{Gompertz.TransLon:115}. Also here, linearization cannot produce any valid mechanisms.

The Gompertz (1825)\nocite{Gompertz.TransLon:115} model remains despite its obviously unrealistic growth rate assumptions still widely in use, cf. Wang and Guo (2024)\nocite{Wang.BioTechAdv:72} and references therein. Our mission is to test under what circumstances commonly used model selection methods (Burnham and Anderson (2002)\nocite{Burnham.Modelselect}) rank this model above other models independently of its mechanistic flaws.

We start again by getting back to the Beverton-Holt (1957)\nocite{Beverton.Dynamics} model (\ref{diskret_logistic})
\begin{displaymath}
  X(T)=F(X_0)=\frac{KX_0}{K\exp(-RT)+(1-\exp(-RT))X_0}
\end{displaymath}
and we introduce the new variables $Y_0=\log X_0$ and $Y(T)=\log Y_0$. We get that
\begin{eqnarray*}
  Y(T)&=&\log F(X_0)=\log\left(\frac{KX_0}{K\exp(-RT)+(1-\exp(-RT))X_0}\right)\\
  &=&\log K+\log X_0-\log\left(K\exp(-RT)+(1-\exp(-RT))X_0\right)\\
  &=&\log K+Y_0-\log\left(K\exp(-RT)+(1-\exp(-RT))\exp(Y_0)\right)\\
  &=&G(Y_0)
\end{eqnarray*}
is a model corresponding to (\ref{diskret_logistic}) on a logarithmic scale. We get that
\begin{displaymath}
G(\log K)=\log K
\end{displaymath}
and that
\begin{displaymath}
  G^\prime(Y_0)=1-\frac{(1-\exp(-RT))\exp(Y_0)}{K\exp(-RT)+(1-\exp(-RT))\exp(Y_0)}.
\end{displaymath}
Therefore, we have
\begin{displaymath}
  G^\prime(\log K)=1-1+\exp(-RT)=\exp(-RT)
\end{displaymath}
and a log-scale linearization is given by
\begin{equation}
  Y(T)=\log K+\exp(-RT)(Y_0-\log K).
\label{discrete_log_linear_model}
\end{equation}
Switching back to the original coordinates, we get that this linearization is the Gompertz (1825)\nocite{Gompertz.TransLon:115} model
\begin{eqnarray}
  X(T)&=&\exp\left(\log K+\exp(-RT)(Y_0-\log K)\right)\nonumber\\
  &=&\exp(\log K)\cdot\exp\left(\frac{Y_0\exp(-RT)}{\log K\cdot\exp(-RT)}\right)\nonumber\\
  &=&K\left(\frac{\exp Y_0}{K}\right)^{\exp(-RT)}=K\left(\frac{X_0}{K}\right)^{\exp(-RT)}\label{diskret_Gompertz}\\
  &=&K^{1-\exp(-RT)}X_0^{\exp(-RT)}.\nonumber
\end{eqnarray}
We note that the species produces unlimited per-capita growth rates at small population densities. This implies that the population must start producing its own limiting resources when scarce. This property has even more severe mechanistical consequences for ecologies involving several species. As an approximation of a predator-prey model, the Gompertz (1825)\nocite{Gompertz.TransLon:115} model predicts that a predator can produce its own prey at low densities!

\section{A differential equation for the log-linear model}
\label{log_linear_de}

The linear model (\ref{Poinc_lin}) given by
\begin{displaymath}
X(T)=\frac{a}{b}(\exp(bT)-1)+\exp(bT)X_0
\end{displaymath}
was derived as a Poincar\'{e} map of the linear differential equation
\begin{displaymath}
  \dot{X}=a+bX.
\end{displaymath}
Comparing the coefficients with (\ref{discrete_log_linear_model}) gives the system of equations
\begin{eqnarray*}
  \frac{a}{b}\left(\exp(bT)-1\right)&=&\log K\left(1-\exp(-RT)\right), \\
   b&=&-R,
\end{eqnarray*}
and we have the solution
\begin{eqnarray*}
  a &=& R\log K, \\
  b &=& -R,
\end{eqnarray*}
by back-substitution. The differential equation producing the Poincar\'{e} map (\ref{discrete_log_linear_model}) is therefore
\begin{displaymath}
  \dot{Y}=R\log{K}-RY
\end{displaymath}
which translates to the Gompertz (1825)\nocite{Gompertz.TransLon:115} model
\begin{equation}
  \dot{X}=RX\log\left(\frac{K}{X}\right)
  \label{Gompertz_de}
\end{equation}
using the transformation $X=\exp(Y)$. In this differential equation case, we also see that the per-capita growth rate is given by
\begin{displaymath}
  R\log\left(\frac{K}{X}\right)
\end{displaymath}
and that
\begin{displaymath}
  \lim_{X\rightarrow 0+}R\log\left(\frac{K}{X}\right)=+\infty
\end{displaymath}
meaning that the model predicts unbounded self-amplification from arbitrary small populations without any external resource input. Since our chemostat is nutrient limited, this is impossible. Similar problems can be confirmed for several extensions of the Gompertz (1825)\nocite{Gompertz.TransLon:115} model. For instance, in the predator-prey extension, the Gompertz functional form implies infinite prey consumption as the predator density tends to zero. We conclude that (\ref{diskret_Gompertz}) shares many of its problems with (\ref{Gompertz_de}). Our objective is to analyze whether census data can be used for rejecting such unrealistic assumptions.

\section{Stochastic models for linear immigration and log-linear models}
\label{Stochastic_models}

Our next step is to create stochastic population growth models corresponding to the local linear immigration model (\ref{lin_de}) that we now express as
\begin{equation}
  \dot{X}=R(K-X)=RK-\rho X+DX
  \label{lin_de_chemo}
\end{equation}
and the Gompertz log-linear differential equation (\ref{Gompertz_de}). We express it as
\begin{equation}
  \dot{X}=RX\log\left(\frac{K}{X}\right)=RX(\log K-\log X)=RX\log K-RX\log X.\label{Gompertz_de_chemo}
\end{equation}
The objective is to compute the likelihood of the models (\ref{lin_de_chemo}) and (\ref{Gompertz_de_chemo}) with respect to data generated by (\ref{logistic_stok}).
A stochastic model corresponding to the linear model (\ref{lin_de_chemo}) is given by
\begin{eqnarray}
\dot{p}_X&=&RKp_{X-1}-RKp_{X}+\rho(X+1)p_{X+1}-\rho Xp_X+D(X-1)p_{X-1}\nonumber\\
&&-DXp_X, \: X=1,2,3,\dots,\label{linear_stok}\\
\dot{p}_0&=&-RKp_{0}+\rho p_1.\nonumber
\end{eqnarray}
We note that it is not possible to find any realistic interpretation of the parameters appearing in this model. They have simply nothing to do with the parameters that appear in the chemostat. The model also differs from (\ref{logistic_stok}) in the sense that the state of extinction is not absorbing.

The problems with the parameters, scaling, and the extinction state are of a considerably larger magnitude for the stochastic model corresponding to Gompertz (1825) differential equation (\ref{Gompertz_de_chemo}). We get formally in this case
\begin{eqnarray}
\dot{p}_X&=&R\log K(X-1)p_{X-1}-R\log K Xp_X\nonumber\\
&&+R(X+1)\log(X+1)p_{X+1}-RX\log Xp_X, \: X=1,2,3,\dots.\label{Gompertz_stok}
\end{eqnarray}
For this stochastic model, it is impossible to reach the extinction state, that is $p_0=0$. We note that one of the important features of stochastic population models is their ability to cover the important phenomenon of population extinction. We note that there is no obvious way for splitting the term $X\log X$ into clearly defined birth- and death-processes. This causes a difference in the number of independent parameters in the linear models (\ref{linear_stok}) and (\ref{Gompertz_stok}). We continue the discussion about the number of independent parameters in our models in Section \ref{detect_nonlin}. Such a splitting is not visible in any of the deterministic approximations, but it is visible in the diffusion approximations.

\begin{figure}
\epsfxsize=170mm
\begin{picture}(264,200)(0,0)
\put(-52,0){\epsfbox{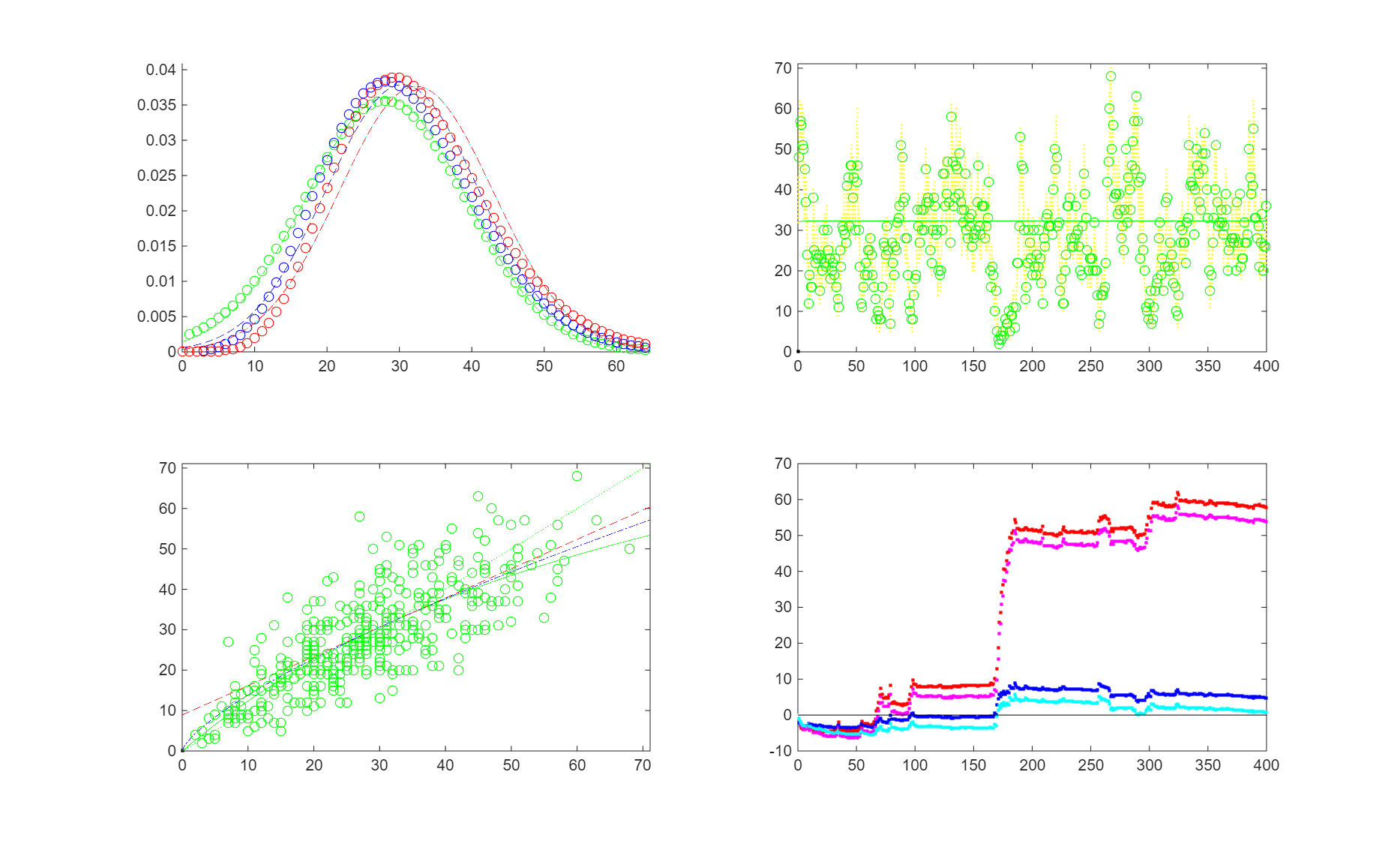}}

\put(158,150){$X$}
\put(-10,275){$p$, $q$}
\put(84,150){(a)}

\put(370,150){$T$}
\put(210,270){$X$}
\put(300,150){(b)}

\put(150,15){$X(T)$}
\put(-3,135){$X(T+1)$}
\put(84,15){(c)}

\put(370,15){$T$}
\put(195,135){$\Delta AIC,\:\Delta BIC$}
\put(300,15){(d)}
\end{picture}
\caption{Parameter values are given by $M=3\cdot 10^{-22}$, $A=.01$, $I_0\approx 2.995\cdot 10^{23}$, $D=.8$, and $L=1$. (a) Quasi-stationary and stationary probability mass function of (\protect\ref{logistic_stok}), (\protect\ref{linear_stok}), and (\protect\ref{Gompertz_stok}), marked with green, red, and blue $\circ$-marks, respectively, together with normal distributions sharing their mean and variance (dashed, same colors) and normal distributions derived as diffusion approximations (dotted, magenta for (\protect\ref{logistic_stok}), (\protect\ref{linear_stok})) and cyan blue for (\protect\ref{Gompertz_stok}). (b) Simulation of (\protect\ref{logistic_stok}) denoted with a yellow curve, census data denoted by green $\circ$-marks. (c) Census data from (\protect\ref{logistic_stok}) denoted by green $\circ$-marks compared to the maps (\protect\ref{diskret_logistic}), (\protect\ref{diskret_linear}), and (\protect\ref{diskret_Gompertz}) (d) Information criteria -differences from the model that generated the data (\protect\ref{logistic_stok}). Red dots denote AIC-differences and magenta dots denote BIC-differences between (\protect\ref{linear_stok}) and (\protect\ref{logistic_stok}) versus the available amount of data. Blue dots denote AIC-differences and cyan dots denote BIC-differences between (\protect\ref{Gompertz_stok}) and (\protect\ref{logistic_stok}) versus the available amount of data.}
\label{likelihoods3}
\end{figure}

We compute the associated variance with respect to the splitting into birth- and death processes that has been made above. The diffusion approximations of the stationary solutions of the models (\ref{linear_stok}) and (\ref{Gompertz_stok}) share the mean
\begin{displaymath}
  K=\frac{M\frac{A}{L}\frac{I_0}{D}-D}{\frac{A}{L}}
\end{displaymath}
and we conclude through our computations in Section \ref{appr_norm_sec} that the variances of the diffusion approximations of the stationary solutions of (\ref{linear_stok}) and (\ref{Gompertz_stok}) are
\begin{equation}
  V_1=-\frac{R^2
  +\left(\rho+D\right)R}{2\alpha R}=-\frac{R\left(\rho-D+\rho+D\right)}{2\alpha R}=\frac{\rho}{\alpha}=\frac{MI_0}{D}\label{Variance_linear}
\end{equation}
and
\begin{eqnarray}
V_2&=&-\frac{RK\log K+RK\log K}{2\left.R(\log K-\log X-1)\right|_{X=K}}\nonumber=\frac{2RK\log K}{2R}=K\log K=\\
&=&\left(\frac{M\frac{A}{L}\frac{I_0}{D}-D}{\frac{A}{L}}\right)\log \left(\frac{M\frac{A}{L}\frac{I_0}{D}-D}{\frac{A}{L}}\right),\label{Variance_Gompertz}
\end{eqnarray}
respectively. In Figure \ref{likelihoods3}(a) we used the parameter values $M=3\cdot 10^{-22}$, $A=.01$, $I_0=2.995\cdot 10^{23}$, $D=.8$, and $L=1$ and plotted the quasi-stationary probability mass functions of (\ref{logistic_stok}) with green $\circ$-marks, the stationary probability mass function of (\ref{linear_stok}) with red $\circ$-marks and the stationary probability mass function of (\ref{Gompertz_stok}) with blue $\circ$-marks. Dashed green, red, and blue curves denote normal distributions that share the means and variances of these probability mass functions. The coinciding diffusion approximations of (\ref{logistic_stok}) and (\ref{linear_stok}) are denoted by dotted curves in magenta whereas the diffusion approximation of (\ref{Gompertz_stok}) is denoted by a dotted curve in cyan blue.

\begin{figure}
\epsfxsize=170mm
\begin{picture}(264,200)(0,0)
\put(-52,0){\epsfbox{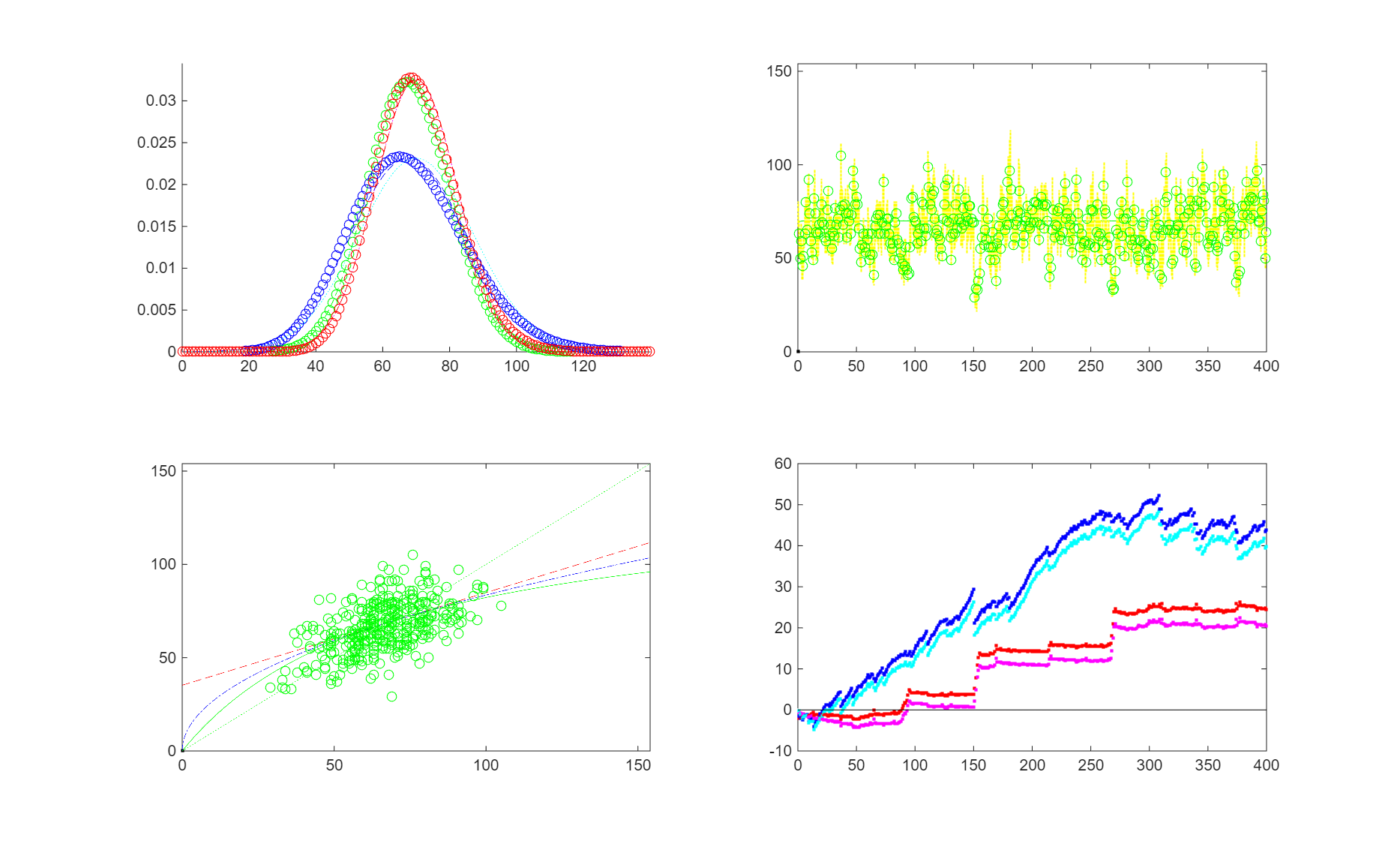}}

\put(158,150){$X$}
\put(-10,270){$p$, $q$}
\put(84,150){(a)}

\put(370,150){$T$}
\put(210,270){$X$}
\put(300,150){(b)}

\put(150,15){$X(T)$}
\put(-3,135){$X(T+1)$}
\put(84,15){(c)}

\put(370,15){$T$}
\put(195,135){$\Delta AIC,\:\Delta BIC$}
\put(300,15){(d)}
\end{picture}
\caption{Parameter values are given by $M=3\cdot 10^{-22}$, $A=.01$, $I_0=4\cdot 10^{23}$, $D=.8$, and $L=1$. (a) Quasi-stationary and stationary probability mass function of (\protect\ref{logistic_stok}), (\protect\ref{linear_stok}), and (\protect\ref{Gompertz_stok}), marked with green, red, and blue $\circ$-marks, respectively, together with normal distributions sharing their mean and variance (dashed, same colors) and normal distributions derived as diffusion approximations (dotted, magenta for (\protect\ref{logistic_stok}), (\protect\ref{linear_stok})) and cyan blue for (\protect\ref{Gompertz_stok}). (b) Simulation of (\protect\ref{logistic_stok}) denoted with a yellow curve, census data denoted by green $\circ$-marks. (c) Census data from (\protect\ref{logistic_stok}) denoted by green $\circ$-marks compared to the maps (\protect\ref{diskret_logistic}), (\protect\ref{diskret_linear}), and (\protect\ref{diskret_Gompertz}) (d) Information criteria -differences from the model that generated the data (\protect\ref{logistic_stok}). Red dots denote AIC-differences and magenta dots denote BIC-differences between (\protect\ref{linear_stok}) and (\protect\ref{logistic_stok}) versus the available amount of data. Blue dots denote AIC-differences and cyan dots denote BIC-differences between (\protect\ref{Gompertz_stok}) and (\protect\ref{logistic_stok}) versus the available amount of data.}
\label{likelihoods4}
\end{figure}
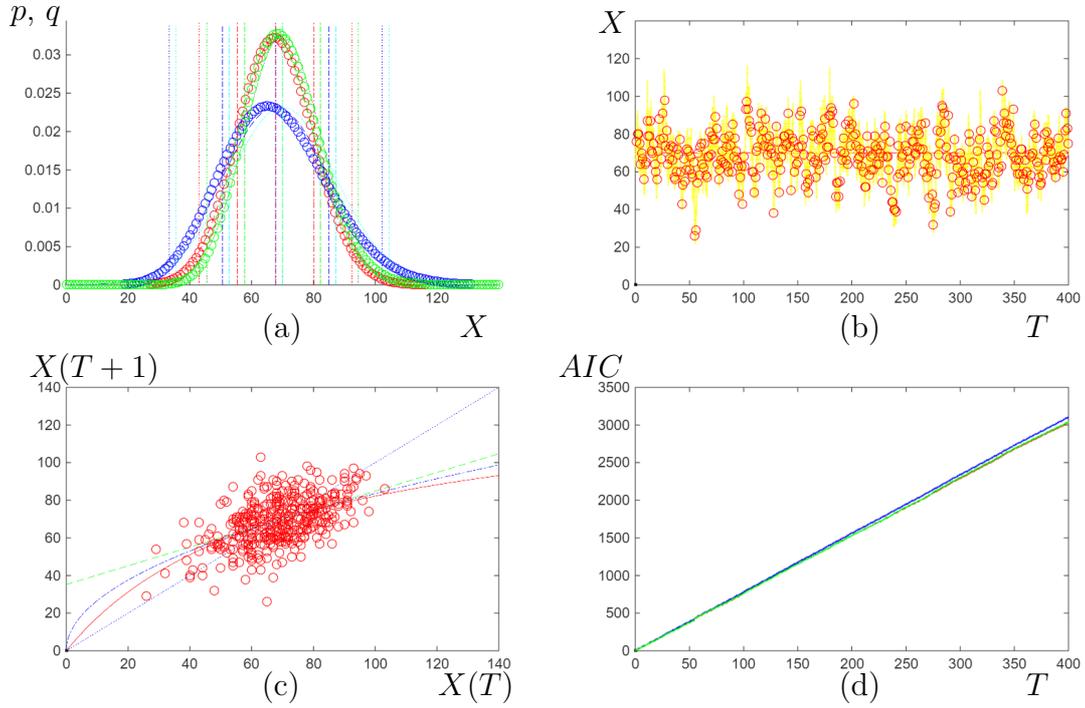

In Figure \ref{likelihoods4}(a) we used the parameter values $M=3\cdot 10^{-22}$, $A=.01$, $I_0=4\cdot 10^{23}$, $D=.8$, and $L=1$ and plotted the quasi-stationary probability mass functions of (\ref{logistic_stok}) with green $\circ$-marks, the stationary probability mass function of (\ref{linear_stok}) with red $\circ$-marks and the stationary probability mass function of (\ref{Gompertz_stok}) with blue $\circ$-marks. Dashed green, red, and blue curves denote normal distributions that share the means and variances of these probability mass functions. The coinciding diffusion approximations of (\ref{logistic_stok}) and (\ref{linear_stok}) are denoted by dotted curves in magenta whereas the diffusion approximation of (\ref{Gompertz_stok}) is denoted by a dotted curve in cyan blue.

\begin{figure}
\epsfxsize=121mm
\begin{picture}(264,200)(0,0)
\put(40,5){\epsfbox{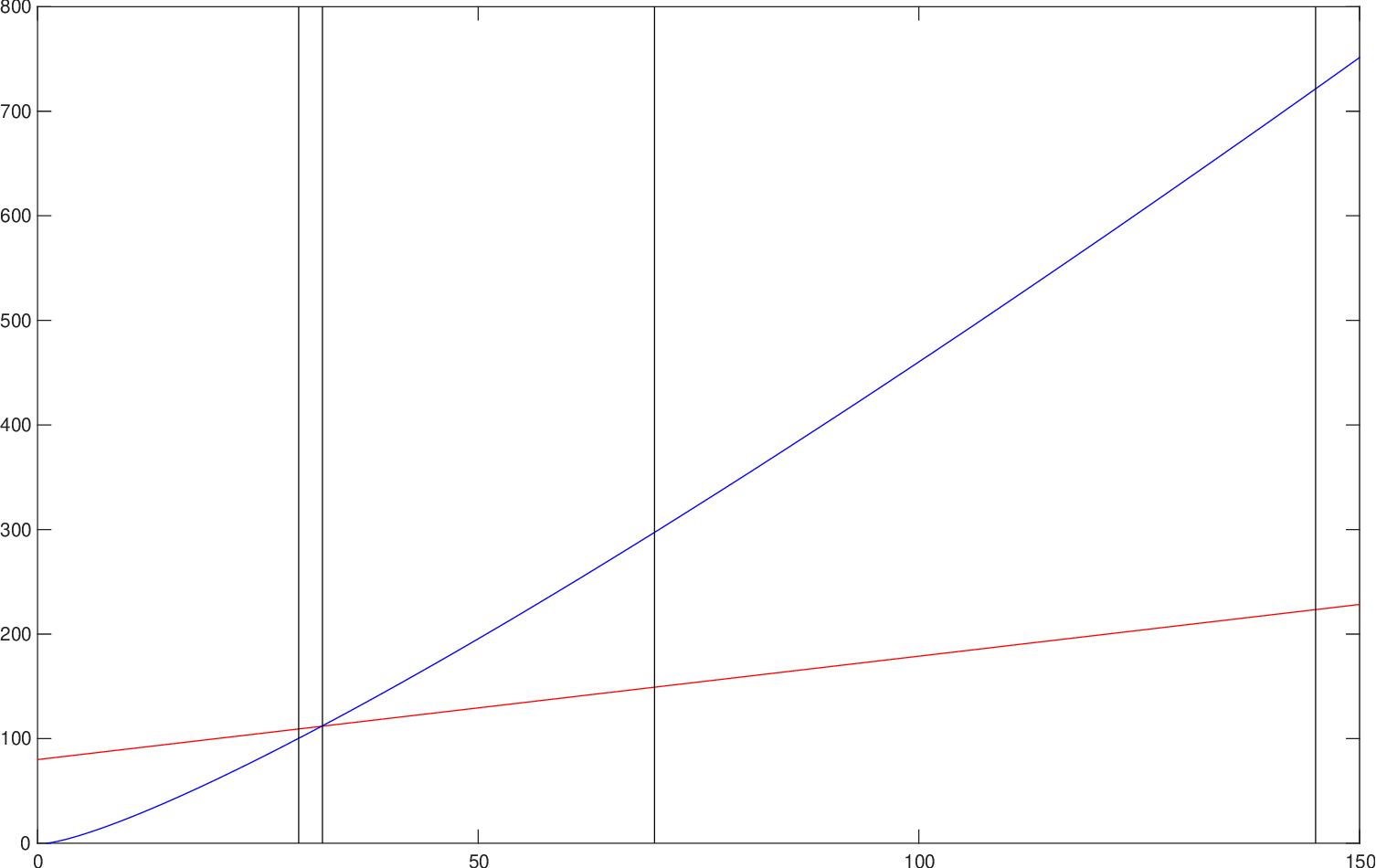}}
\put(38,202){$V$}
\put(363,0){$K$}
\put(106.5,0){(a)}
\put(112,224){(b)}
\put(197,224){(c)}
\put(361,224){(d)}
\put(220,175){$V_2=K\log K$}
\put(245,170){\vector(2,-3){20}}
\put(260,100){$V_1=\frac{LD}{A}+K$}
\put(280,95){\vector(1,-3){10}}
\put(10,29){$\frac{LD}{A}$}
\put(30,32){\vector(1,0){15}}
\end{picture}
\caption{The variances (\protect\ref{Variance_linear}) and (\protect\ref{Variance_Gompertz}) of the diffusion approximations of the stationary distributions as functions of the carrying capacity of the models (\protect\ref{linear_stok}) and (\protect\ref{Gompertz_stok}) denoted as a red line and  denoted as a blue curve, respectively. The parameter values are $M=3\cdot 10^{-22}$, $A=.1$, $D=.8$, and $L=1$. The black vertical lines correspond to (a) $I_0=I_0^\ast(0)\approx 2.923 \cdot10^{23}$, (b) $I_0\approx 2.995\cdot10^{23}$, (c) $I_0=4\cdot10^{23}$, and (d) $I_0=6\cdot10^{23}$. The case (b) corresponds to the case where the diffusion approximations of the quasi- and stationary distributions of all three models (\protect\ref{logistic_stok}), (\protect\ref{linear_stok}), and (\protect\ref{Gompertz_stok}) share variances.}
\label{Variancs_comparison}
\end{figure}
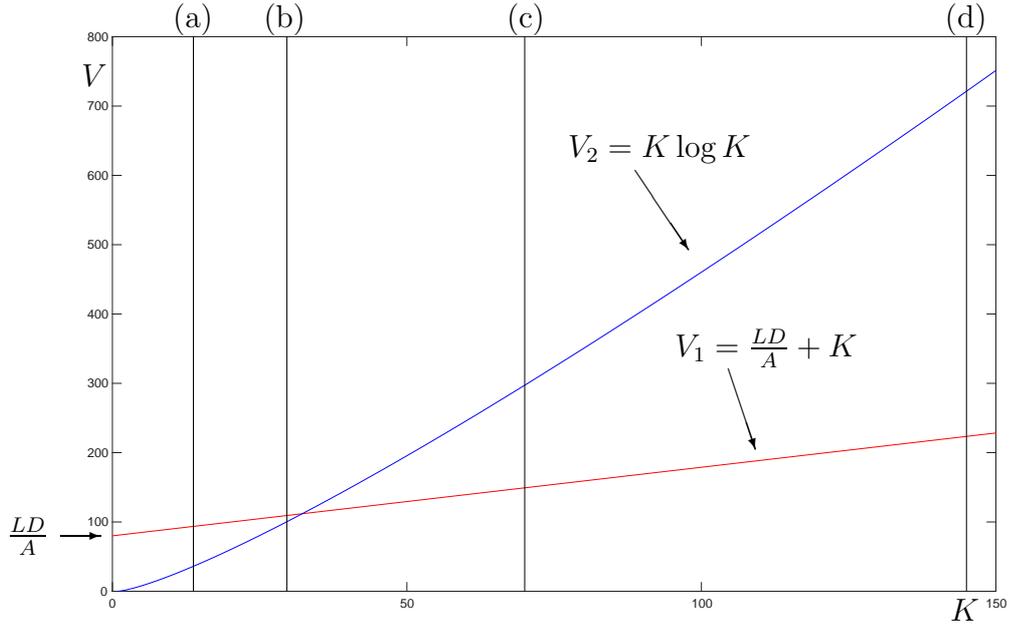

We conclude from (\ref{Variance_linear}) and (\ref{Variance_Gompertz}), that the two variances, $V_1$ and $V_2$, can be expressed as functions of the carrying capacities (cf. (\ref{KdepVar})) according to
\begin{displaymath}
V_1(K)=\frac{LD}{A}+K\:{\rm{and}}\:V_2(K)=K\log K.
\end{displaymath}
We have $V_2(1)=0<V_1(1)$ and $V_2(K)>V_1(K)$ for a sufficiently high $K$. We also have
\begin{displaymath}
  V_2^\prime(K)-V_1^\prime(K)=\log K+1-1=\log K\geq 0, \:{\rm{if}}\: K\geq 1.
\end{displaymath}
Therefore, we have $V_2>V_1$ as soon as the quasi-stationary distribution starts becoming normal. The situation is depicted in Figure \ref{Variancs_comparison}. Here we have denoted the variance $V_1$ of the stationary distribution of the diffusion approximation of model (\ref{linear_stok}) with red line. The variance $V_2$ of the stationary distribution of the diffusion approximation of model (\ref{Gompertz_stok}) with is denoted with a blue curve.

A comparison of Figure \ref{likelihoods3}(a) and Figure \ref{likelihoods4}(a) confirms that the variance of the model (\ref{Gompertz_stok}) becomes significantly larger than the variance of (\ref{logistic_stok}) and (\ref{linear_stok}) as the deterministic carrying capacity increases and we approach normality. In Figure \ref{likelihoods3}, we have selected the parameters so that the variances of the diffusion approximations of all three models (\ref{logistic_stok}), (\ref{linear_stok}), and (\ref{Gompertz_stok}) receive an equal value.

\section{Detecting nonlinearities}
\label{detect_nonlin}

Our next question is whether the nonlinearity in the model (\ref{diskret_logistic}) can be detected from census data generated by the Markov process (\ref{logistic_stok}). We start using the parameter values $M=3\cdot 10^{-22}$, $A=.01$, $I_0\approx 2.995\cdot 10^{23}$, $D=.8$, and $L=1$ and one resulting outcome starting from $X_0=32$ is depicted with a yellow curve in Figure \ref{likelihoods3}(b).

Simulations of such outcomes can be done in an efficient manner by e. g the Gillespie's (1976,1977)\nocite{Gillespie_JoCP:22,Gillespie.JoPC:81} direct algorithm. It starts with the generation of two independent identically distributed random numbers on the unit interval. The first random number is used for computing the time to the next event by assuming an exponentially distributed inter-event time. The second random number is used for deciding the type of event and they are of three types corresponding to the six terms in the first equation of (\ref{logistic_stok}): (a) linear births corresponding to terms 1-2, (b) linear deaths corresponding to terms 3-4, and (3) quadratic deaths to the terms 5-6. There are sources that contain sample code for this algorithm (for MatLab code see e.g. Allen (2008)\nocite{Brauer_Driesche_cpt3}, p125-126) and they can easily be modified to cover our situation.

Outcomes at the times $T=1,\cdots,400$ are denoted by green $\circ$-marks and we use $X_0$, $X_1$,\dots $X_{400}$ to denote these outcomes below. They are used as our simulated ecological data for the nonlinear model (\ref{diskret_logistic}). The only stochastic part appearing in the data is from the model (\ref{logistic_stok}) and models the existence of a limited number of individuals in the population, cf. Bailey (1964)\nocite{Bailey.Elements}. 

We now compare our simulated data to the models (\ref{diskret_logistic}), (\ref{diskret_linear}) and (\ref{diskret_Gompertz}) to the identity line denoted as a dotted green line (cobwebbing, cf. Devaney (2022)\nocite{devaneybokIII}) in Figure \ref{likelihoods3}(c) and these models are denoted by curves in solid green, dashed red, and dashdotted blue, respectively. Note that the deterministic part of the log-linear model (\ref{diskret_Gompertz}) is significantly closer to (\ref{diskret_logistic}) than (\ref{diskret_linear}). Despite the linear properties of the model, it describes a part of the nonlinearity of (\ref{diskret_logistic}).

In Figure \ref{likelihoods4}(b)-(c) we repeat the steps in the previous paragraphs for the parameter values $M=3\cdot 10^{-22}$, $A=.01$, $I_0=4\cdot 10^{23}$, $D=.8$, and $L=1$. The outcome that is depicted by a yellow curve in Figure \ref{likelihoods4}(b) starts in this case from $X_0=69$. It is evident that the outcomes marked by green $\circ$-marks remain far from extinction and the intercepts of the models as we get closer to the normal distribution.

The next step is to apply model selection criterion to determine how close the various models are with respect to the data and how many parameters can be justified in (\ref{diskret_logistic}). Two commonly used model-selection criteria with many good properties are the Akaike (1973)\nocite{Akaike.second} and Bayesian information criteria (AIC and BIC), see e.g. Burnham and Anderson (2002)\nocite{Burnham.Modelselect}. These two criteria are defined according to
\begin{displaymath}
  {\rm{AIC}}=-2\log({\cal{L}({\rm{model}}|{\rm{data}})})+2P
\end{displaymath}
and
\begin{displaymath}
  {\rm{BIC}}=-2\log({\cal{L}({\rm{model}}|{\rm{data}})})+P\log n
\end{displaymath}
where $-2\log({\cal{L}({\rm{model}}|{\rm{data}})})$ is the likelihood of the model given the set of data, $P$ is the number of parameters in the model including the intercept and the variance, and $n$ is the sample size of the data. Both criteria contain the likelihood of the model and a term penalizing the number of parameters in the model.

The number of independent parameters in the model is a quantity that deserves some discussion. This is most visible from the equations for the birth-death processes (\ref{logistic_stok}), (\ref{linear_stok}) and (\ref{Gompertz_stok}) that we used here. This reveals that independent parameter combinations in (\ref{logistic_stok}) and (\ref{linear_stok}) are $\rho$, $D$, and $\alpha$ but for (\ref{Gompertz_stok}) they are $R$ and $K$. The smaller number of independent parameters in (\ref{Gompertz_stok}) was due the lack of possibilities to specify valid birth and death processes for (\ref{Gompertz_stok}) in any appropriate manner. For all data any quantities like slopes, intercepts, and variances must be expressed in terms of the parameters mentioned above.

However, use of the information that (\ref{logistic_stok}) contains just three independent parameters would be to apply a procedure that assumes that we know the type of nonlinearity we are looking for. Our purpose was to clarify whether model selection methods can detect nonlinearities with respect to typical ecological census data. This means that the linear models have the slope, the intercept, and the variance as independent parameters whereas a nonlinear model possesses an additional justified parameter describing some part of the nonlinearity. This would imply deriving a stochastic model from (\ref{quadr_model}) without the pre-assumption that any of its parameters can be put equal to zero.

Therefore, we have $P=3$ for the linear models (\ref{linear_stok})-(\ref{Gompertz_stok}) and $P=4$ for the nonlinear model (\ref{logistic_stok}) that generated our data. Strictly speaking, we could have used $P=2$ for (\ref{Gompertz_stok}). However, the main message here is that there are conditions where (\ref{Gompertz_stok}) can be justified as some kind of description of data from (\ref{logistic_stok}). Such a window will be present already when we use $P=3$ for (\ref{Gompertz_stok}). Strictly speaking, we have at least $P=4$ for the nonlinear model, too, since there might be reasons to split each of the deterministic terms in (\ref{quadr_model}) into different birth and death-terms in its stochastic description, too. The validation of a nonlinear model in ecology might thus be harder than inferred here.

Assume now that we use the model (\ref{logistic_stok}) to generate the data set
\begin{displaymath}
X_0,X_1,\dots,X_n.
\end{displaymath}
Then $X_i$, $i=1,\dots,n$ are integers corresponding to the population abundances at the discrete time-instances $t=1,\dots,n$. We use
\begin{displaymath}
  p_X(0)=\left\{\begin{array}{cc}
  1,& X=X_0\\
  0,& X\neq X_0
  \end{array}\right.
\end{displaymath}
as an initial probability mass function for the model (\ref{logistic_stok}) and compute the probability $p_{X_1}(1)$ given that $p_{X_0}(0)=1$. Let us denote this probability by $p_{X_1}(1)|_{p_{X_0}(0)=1}$ to make this conditioning clear. The likelihood of the model (\ref{logistic_stok}) that generated the data is then given by
\begin{displaymath}
  {\cal{L}}({\rm{(\ref{logistic_stok})}}|X_0\dots X_n)
  =\Pi_{i=1}^n p_{X_i}(1)|_{p_{X_{i-1}}(0)=1}.
\end{displaymath}
The computation of the logarithm of this quantity is less subject to underflow than the computation of this quantity itself since most of these probabilities are very small and so are their products. Similar formulae hold for the likelihoods of the models (\ref{linear_stok}) and (\ref{Gompertz_stok}). Note that the set of data remains unchanged, but we compute the required conditional probabilities using the models (\ref{linear_stok}) and (\ref{Gompertz_stok}) instead. There exist ways to compute the likelihoods with the aid of the normal approximations, but they are not used here. Our computation of the likelihood is therefore not sensitive to the accuracy of the normal approximations.

In Figure \ref{likelihoods3}(d) we plot the AIC differences between the models (\ref{logistic_stok}) and (\ref{linear_stok}) (red dots), and (\ref{logistic_stok}) and  (\ref{Gompertz_stok}) (blue dots), respectively. The corresponding BIC differences are denoted by magenta dots and cyan dots, respectively. We still use the parameter values $M=3\cdot 10^{-22}$, $A=.01$, $I_0=3\cdot 10^{23}$, $D=.8$, and $L=1$. The AIC-criterion ranks the model with the lowest AIC highest, and BIC-criterion works similarly. According to Figure \ref{likelihoods3}(a) all three models (\ref{logistic_stok}), (\ref{linear_stok}), and (\ref{Gompertz_stok}) possess approximately the same means and variances. Since the means of all models are quite low, the differences in the deterministic models near the origin play a role when distinguishing between the models. The AIC value of the linear model (\ref{linear_stok}) becomes higher than the corresponding value for the logistic model (\ref{logistic_stok}) and the log-linear model (\ref{Gompertz_stok}) already for approximately $70\dots 80$ data points. We note that this difference becomes very high after that the cluster of points near the origin is added to the data approximately at $170\dots 180$ data points. The difference between the log-linear (\ref{Gompertz_stok}) and the logistic model (\ref{logistic_stok}) become somewhat visible after the cluster of points a have $170\dots 180$ data points, too, but this difference is not large enough to remain persistent over time. Since the differences reported here are dependent on clusters of points that correspond to unusually low population values, different samples may end up with quite different results.

In Figure \ref{likelihoods4}(d) we plotted a similar plot as in Figure \ref{likelihoods3}, but the parameter values are now $M=3\cdot 10^{-22}$, $A=.01$, $I_0=4\cdot 10^{23}$, $D=.8$, and $L=1$. According to Figure \ref{likelihoods4}(a) the models (\ref{logistic_stok}) and (\ref{linear_stok}) share their variances whereas the model (\ref{Gompertz_stok}) possesses a significantly higher variance.

We conclude that the log linear model (\ref{Gompertz_stok}) is rejected based on very little data (approximately $50\dots 60$ data points) despite that its deterministic part (\ref{diskret_Gompertz}) is closer to the deterministic part of the correct model (\ref{diskret_logistic}) than the deterministic part of the linear model (\ref{diskret_linear}). The rejection of (\ref{Gompertz_stok}) and its deterministic part (\ref{diskret_Gompertz}) is therefore, based on differences in the variances and (\ref{Variance_linear}) and (\ref{Variance_Gompertz}).

Since the variances and means of the models (\ref{logistic_stok}) and (\ref{linear_stok})  coincide and correspond to approximately normal distributions, the data remains sufficiently close to the carrying capacity. This requires huge data sets for detection of any nonlinearities. Also here, we note that clusters of data corresponding to exceptionally small populations sizes start helping to reject the model (\ref{linear_stok}) after data levels of $155\dots 160$ and $270\dots 280$. This kind of exceptional data that can be used for distinguishing between the models is expected to become rarer as normal approximations become better for higher carrying capacities. Collection of the corresponding amounts of data can be associated with huge costs in ecology. We note the length of the commonly re-analyzed classical ecological time-series by Nicholson (1954)\nocite{Nicholson.AustJZool:2} and Utida (1957)\nocite{Utida.ColdSpring:22} share the order of a few hundred data points.

\section{Summary}
\label{sum}

In this chapter we have studied a species growing in chemostat conditions obeying the stochastic logistic growth equation. The properties of the stochastic logistic equation are well-known. The extinction state is absorbing, and it does not possess any stationary distribution. Nevertheless, a quasi-stationary distribution exists, if we condition on non-extinction.

If the carrying capacity is sufficiently large, then the quasi-stationary
solution is approximately normal. Since tails of normal distributions possess very low probabilities the expected lifetimes of such populations are
very long (N{\aa}sell (2001)\nocite{Naasell.JTB:211}). Therefore, the general rule is that measures against invasive species and diseases must be taken as soon as they are detected or encountered (Palmer, Whalen, Aley, Russell (2026)\nocite{Palmer.BCons:316}).
The stochastic logistic equation has therefore the potential to describe phenomena that are well in concordance with experience.

The parameters defining the birth- and death-processes are, however, seldom
coupled to the resource dynamics of the system. Here, we associate the birth- and death processes to the chemostat conditions. This couples the parameters
appearing in the stochastic logistic equation to the resource dynamics.

Nonlinear models can usually be justified by mechanistic arguments. Nevertheless, linear models on various scales may be able to predict collected census data equally well.

We linearize the stochastic logistic model that we derived from the
chemostat both without scaling and on a logarithmic scale. Without scaling,
the linearized model describes immigration and linear growth. The
log-scaled linear model has no mechanistic interpretation at all.

For high carrying capacities corresponding to nearly normally distributed
data, we conclude that the stochastic logistic model and the linear model
describing immigration and linear growth describe the data equally well. The
reason is that the quasi-stationary distribution of the stochastic logistic
model shares its mean and variance with the stationary distribution of the
linearized model. Clusters of outliers describing the dynamics at unusually low population densities are required for distinguishing these models from each other.

For low carrying capacities that correspond to high probabilities of quite low
numbers of individuals in the quasi-stationary distribution of the stochastic
logistic model, the model that was linear on logarithmic scale gives a better
description of the data than the model corresponding to linear growth
combined with immigration. The reason is matching intercepts, the existence of data near the intercept, and matching variances. Also in this case, existence of data describing the dynamics at low densities is required for distinguishing between the models. In this case, however, we stay farther away from normal approximations, and such data is consequently expected to be more common.


\bibliographystyle{abbrv}
\bibliography{artiklar,biologi,dynamic}

\begin{thebibliography}{10}

\bibitem{Akaike.second}
H.~Akaike.
\newblock Information theory and an extension of the maximum likelihood
  principle.
\newblock In B.~N. Petrov and F.~Csaki, editors, {\em 2nd International
  Symposium on Information Theory}, pages 267--281. Akademia Kiado, Budapest,
  1973.

\bibitem{Brauer_Driesche_cpt3}
L.~J.~S. Allen.
\newblock An introduction to stochastic epidemic models.
\newblock In F.~Brauer, P.~van~den Driessche, and J.~Wu, editors, {\em
  Mathematical Epidemiology}, volume 1945 of {\em Lecture Notes in
  Mathematics}. Springer, Berlin, 2008.

\bibitem{Bailey.Elements}
N.~T.~J. Bailey.
\newblock {\em The Elements of Stochastic Processes}.
\newblock Wiley, New York, 1964.

\bibitem{Beverton.Dynamics}
R.~J.~H. Beverton and S.~J. Holt.
\newblock {\em On the Dynamics of Exploited Fish Populations}, volume~19 of
  {\em Fisheries Investigation Series 2}.
\newblock Ministry of Agriculture, Fisheries and Food, London, 1957.

\bibitem{BhatMiller}
U.~N. Bhat and G.~K. Miller.
\newblock {\em Elements of Applied Stochastic Processes}.
\newblock John Wiley {\&} Sons, New Jersey, third edition, 2002.

\bibitem{Burnham.Modelselect}
K.~P. Burnham and D.~R. Anderson.
\newblock {\em Model Selection and Multimodel Inference}.
\newblock Springer, 2002.

\bibitem{devaneybokIII}
R.~Devaney.
\newblock {\em An Introduction to Chaotic Dynamical Systems}.
\newblock Addison-Wesley Publishing Company, Inc, third edition, 2022.

\bibitem{Dumortierbook}
F.~Dumortier, J.~Llibre, and J.~C. Art{\'{e}}s.
\newblock {\em Qualitative Theory of Planar Differential Systems}.
\newblock Springer, 2006.

\bibitem{Fokker.AdPh:348}
A.~D. Fokker.
\newblock {D}ie mittlere {E}nergie rotierender elektrischer {D}ipole im
  {S}tralungsfeld.
\newblock {\em {A}nnalen der {P}hysik}, 348:810--820, 1914.

\bibitem{Gillespie_JoCP:22}
D.~T. Gillespie.
\newblock A general method for numerically simulating the stochastic time
  evolution of coupled chemical reactions.
\newblock {\em Journal of Computional Physics}, 22(4):403--434, 1976.

\bibitem{Gillespie.JoPC:81}
D.~T. Gillespie.
\newblock Exact stochastic simulation of coupled chemical reactions.
\newblock {\em The Journal of Physical Chemistry}, 81:2340--2361, 1977.

\bibitem{Gompertz.TransLon:115}
B.~Gompertz.
\newblock On the nature of the function expressive law of human mortality, and
  on the new mode of determining the value of life contingencies.
\newblock {\em Transactions of the Royal Society of London}, 115:513--585,
  1825.

\bibitem{Green.Nonparam}
P.~J. Green and B.~W. Silverman.
\newblock {\em Nonparametric Regression and Generalized Linear Models},
  volume~58 of {\em Monographs on Statistics and Applied Probability}.
\newblock Chapman {\&} Hall, 1994.

\bibitem{Hassel.JAE:45}
M.~P. Hassell, J.~H. Lawton, and R.~M. May.
\newblock Patterns of dynamical behaviour in single-species populations.
\newblock {\em Journal of Animal Ecology}, 45:471--486, 1976.

\bibitem{Hastie}
T.~J. Hastie and R.~J. Tibshirani.
\newblock {\em Generalized Additive Models}.
\newblock Chapman \& Hall, London, 1990.

\bibitem{hirschsmaledevaney}
M.~W. Hirsch, S.~Smale, and R.~L. Devaney.
\newblock {\em Differential Equations, Dynamical Systems, and an Introduction
  to Chaos}.
\newblock Academic Press, Oxford, 2013.

\bibitem{Holling.CanEnt:91}
C.~S. Holling.
\newblock Some characteristics of simple types of predation and parasitism.
\newblock {\em The Canadian Entomologist}, 91(7):385--398, 1959.

\bibitem{jordan}
D.~W. Jordan and P.~Smith.
\newblock {\em Nonlinear Ordinary Differential Equations}.
\newblock Clarendon Press, Oxford, second edition, 1990.

\bibitem{Kooi.BoMB:60}
B.~W. Kooi, M.~P. Boer, and S.~A. L.~M. Kooijman.
\newblock On the use of the logistic equation in models of food chains.
\newblock {\em Bulletin of Mathematical Biology}, 60:231--246, 1998.

\bibitem{Kuehn_Mom_closure}
C.~Kuehn.
\newblock Moment closure - a brief review.
\newblock In E.~Sch{\"{o}}ll, S.~H.~L. Klapp, and P.~H{\"{o}}vel, editors, {\em
  Control of Self-Organizing Nonlinear Systems}, pages 253--271. Springer,
  2016.

\bibitem{Kuno.RPE:33}
E.~Kuno.
\newblock Some strange properties of the logistic equations defined with $r$
  and $k$: Inherent defects or artifacts.
\newblock {\em Researches on Population Ecology}, 33:33--39, 1991.

\bibitem{veszprem1}
T.~Lindstr{\"{o}}m.
\newblock Dependencies between competition and predation - and their
  consequences for initial value sensitivity.
\newblock {\em {SIAM} Journal of Applied Mathematics}, 59(4):1468--1486, 1999.

\bibitem{Lindstr.CSF:42}
T.~Lindstr{\"{o}}m.
\newblock Detecting chaos requires careful analysis of nearly periodic data.
\newblock {\em Chaos, Solitons and Fractals}, 42:212--223, 2009.

\bibitem{lindstr_cheng}
T.~Lindstr{{\"{o}}}m and Y.~Cheng.
\newblock Uniqueness of limit cycles for a limiting case of the chemostat: does
  it justify the use of logistic growth rates.
\newblock {\em Electronic Journal of Qualitative Theory of Differential
  Equations}, 47:1--14, 2015.
\newblock {\tt{http://www.math.u-szeged.hu/ejqtde}}.

\bibitem{Lloyd.TPB:65}
A.~L. Lloyd.
\newblock Estimating variability in models for recurrent epidemics: assessing
  the use of moment closure techniques.
\newblock {\em Theoretical Population Biology}, 65:49--65, 2004.

\bibitem{Marrec.EcolEvol:2}
L.~Marrec, C.~Bank, and T.~Bertrand.
\newblock Solving the stochastic dynamics of population growth.
\newblock {\em Ecology and Evolution}, 13(8):1--20, 2023.

\bibitem{May.Science:186}
R.~M. May.
\newblock Biological {P}opulations with {N}onoverlapping {G}enerations:
  {S}table {P}oints, {S}table {C}ycles, and {C}haos.
\newblock {\em Science}, 186:645--647, 1974.

\bibitem{May.Nature:261}
R.~M. May.
\newblock Simple mathematical models with very complicated dynamics.
\newblock {\em Nature}, 261:459--467, 1976.

\bibitem{Morris.Ecology:71}
W.~F. Morris.
\newblock Problems in detecting chaotic behavior in natural populations by
  fitting simple discrete models.
\newblock {\em Ecology}, 71(5):1849--1862, 1990.

\bibitem{Naasell.JTB:211}
I.~N{\aa}sell.
\newblock Extinction and quasi-stationarity in the {V}erhulst logistic model.
\newblock {\em Journal of Theoretical Biology}, 211(1):11--27, 2001.

\bibitem{Naasell.TPB:64}
I.~N{\aa}sell.
\newblock An extension of the moment-closure method.
\newblock {\em Theoretical Population Biology}, 64:233--239, 2003.

\bibitem{Naasell.BoMB:79}
I.~N{\aa}sell.
\newblock An alternative to moment closure.
\newblock {\em Bulletin of Mathematical Biology}, 79:2088--2108, 2017.

\bibitem{Nicholson.AustJZool:2}
A.~J. Nicholson.
\newblock An outline of the dynamics of animal populations.
\newblock {\em Australian Journal of Zoology}, 2:9--65, 1954.

\bibitem{Nisbet.TPB:40}
R.~M. Nisbet, E.~Mc{C}auley, A.~M. {d}e Roos, W.~W. Murdoch, and W.~S.~C.
  Gurney.
\newblock Population dynamics and element recycling in an aquatic
  plant-herbivore system.
\newblock {\em Theoretical Population Biology}, 40:125--147, 1991.

\bibitem{Palmer.BCons:316}
A.~Palmer, S.~Whalen, J.~P. Aley, and J.~C. Russell.
\newblock Animals or the environment? {P}ublic preference for trade-offs in
  invasive animal management.
\newblock {\em Biological Conservation}, 316:111769, 2026.

\bibitem{perkobok_2001}
L.~Perko.
\newblock {\em Differential Equations and Dynamical Systems}.
\newblock Springer, New York, 2001.

\bibitem{Planck}
M.~Plank.
\newblock {\"{U}}ber einen {S}atz der statistischen {D}ynamik und seine
  {E}rweiterung in der {Q}uantenteorie.
\newblock {\em {S}itzungsberichte der {P}reussischen {A}kademie der
  {W}issenschften zu {B}erlin}, 24, 1917.

\bibitem{renshaw2}
E.~Renshaw.
\newblock {\em Stochastic Population Processes}.
\newblock Oxford University Press, 2011.

\bibitem{Saha.IJDC:11}
P.~Saha and U.~Ghosh.
\newblock Complex dynamics and control analysis of an epidemic model with
  non-monotone incidence and saturated treatment.
\newblock {\em International Journal of Dynamics and Control}, 11:301--323,
  2023.

\bibitem{Schwartz.AS:6}
G.~Schwartz.
\newblock Estimating the dimension of the model.
\newblock {\em The Annals of Statistics}, 6(2):461--464, 1978.

\bibitem{Trostle.Biometrics:80}
P.~Trostle, J.~Guinness, and B.~J. Reich.
\newblock A {G}aussian-process approximation to a spatial {SIR} process using
  moment closures and emulators.
\newblock {\em Biometrics}, 80(3):1--9, 2024.

\bibitem{Utida.ColdSpring:22}
S.~Utida.
\newblock Population fluctuation, an experimental and theoretical approach.
\newblock {\em Cold Spring Harbor Symposia on Quantitative Biology},
  22:139--151, 1957.

\bibitem{verhulst1838}
P.~F. Verhulst.
\newblock Notice sur la loi que la population suit dans son accroisement.
\newblock {\em Corr. Math. et Phys.}, 10:113--121, 1838.

\bibitem{Wang.BioTechAdv:72}
J.~Wang and X.~Guo.
\newblock The {G}ompertz model and its applications in micobial growth and
  bioproduction kinetics: Past, present and future.
\newblock {\em Biotechnology Advances}, 72:108335, 2024.

\bibitem{Whittle.JRSS_B:19}
P.~Whittle.
\newblock On the use of normal approximation in the treatment of stochastic
  processes.
\newblock {\em Journal of the Royal Statistical Society B}, 19:268--281, 1957.

\end{thebibliography}

\appendix

\section{Derivation of the saddle-node threshold}
\label{saddlenodethresholdchapter}

In this appendix we derive the saddle-node threshold (\ref{I0q1}) of the system (\ref{Moment_close_ode}). The isocline corresponding to $\dot{E}[X]=0$ is given by the parabola
\begin{equation}
 V[X]=\frac{E[X]}{\alpha}\left(\rho-D-\alpha E[X]+q_1\left(D+\alpha\right)\right)
 \label{parabola}
\end{equation}
whereas the isocline corresponding to $\dot{V}[X]=0$ is given by the hyperbola
\begin{displaymath}
 V[X]=\frac{E[X]\left(\rho+D+\alpha E[X]+q_1\left(D+\alpha \right)E[X]\right)}{-\alpha+2D+4\alpha E[X]-2\rho-q_1\left(D+\alpha \right)}.
\end{displaymath}
For non-trivial equilibria we must have
\begin{displaymath}
  \frac{\rho-D-\alpha E[X]+q_1\left(D+\alpha\right)}{\alpha}=\frac{\rho+D+\alpha E[X]+q_1\left(D+\alpha \right)E[X]}{-\alpha+2D+4\alpha E[X]-2\rho-q_1\left(D+\alpha \right)}.
\end{displaymath}
Thus, non-trivial equilibria satisfy the quadratic equation
\begin{eqnarray*}
&&\!\!(\rho-D-\alpha E[X]+q_1\left(D+\alpha\right))(-\alpha+2D+4\alpha E[X]-2\rho-q_1\left(D+\alpha \right))\\
&=&-\alpha\rho+2D\rho+4\alpha\rho E[X]-2\rho^2-q_1\rho(D+\alpha)+\alpha D-2D^2-4\alpha DE[X]\\
&&+2\rho D+q_1D(D+\alpha)+\alpha^2E[X]-2D\alpha E[X]-4\alpha^2 E^2[X]+2\rho\alpha E[X]\\
&&+q_1\alpha E[X](D+\alpha)-\alpha q_1(D+\alpha)+2Dq_1(D+\alpha)+4\alpha q_1(D+\alpha) E[X]\\
&&-2\rho q_1(D+\alpha)-q_1^2(D+\alpha)^2\\
&=&\alpha\rho+\alpha D+\alpha^2 E[X]+\alpha q_1\left( D+\alpha \right)E[X]
\end{eqnarray*}
which we write as
\begin{eqnarray*}
&&-4\alpha^2 E^2[X]+\alpha(6R+4q_1(D+\alpha))E[X]\\
&&-2\alpha\rho-2R^2-q_1(D+\alpha)(3R+\alpha+q_1(D+\alpha))=0.
\end{eqnarray*}
Some further simplification gives that
\begin{eqnarray}
&&-2\alpha E^2[X]+(3R+2q_1(D+\alpha))E[X]\nonumber\\
&&-\rho-RK-\frac{q_1}{2}(D+\alpha)\left(3K+1+\frac{q_1}{\alpha}(D+\alpha)\right)=0.
\label{quadr_eq}
\end{eqnarray}
Equation (\ref{quadr_eq}) has real solutions whenever
\begin{eqnarray}
&&(3R+2q_1(D+\alpha))^2\nonumber\\
&&-8\alpha\left(\rho+RK+\frac{q_1}{2}(D+\alpha)\left(3K+1+\frac{q_1}{\alpha}(D+\alpha)\right)\right)\label{quad_R}\\
&=&R^2-8\alpha R-8\alpha D-4q_1(D+\alpha)\alpha\geq 0.\nonumber
\end{eqnarray}
To ensure inequality (\ref{quad_R}) we need
\begin{displaymath}
  R\geq 4\alpha+2\sqrt{4\alpha^2+2\alpha D+q_1\alpha(D+\alpha)}
\end{displaymath}
or
\begin{displaymath}
  M\frac{A}{L}\frac{I_0}{D}\geq 4\frac{A}{L}+D+2\sqrt{4\frac{A^2}{L^2}+2\frac{A}{L}D+q_1\frac{A}{L}(D+\frac{A}{L})}
\end{displaymath}
in the original parameters. This gives the saddle-node threshold in (\ref{I0q1}). The non-trivial equilibria in (\ref{mom_closure_est}) can be derived from (\ref{parabola}) and (\ref{quadr_eq}).
\end{document}